\documentclass[12pt,journal,draftcls,letterpaper,onecolumn]{IEEEtran}
\usepackage{amscd}
\usepackage{mathbbol}
\usepackage{amsfonts,latexsym,subfigure,graphicx,amsmath,algpseudocode,booktabs, balance}
\usepackage[outdir=./picture/]{epstopdf}
\usepackage{algorithm}
\usepackage[usenames,dvipsnames]{pstricks}
\usepackage{epsfig}
\usepackage{pst-grad} 
\usepackage{pst-plot} 
\usepackage{multirow}
\usepackage{amsmath} 
\usepackage{amssymb}  

\newtheorem{defn}{\textsc{Definition}}
\newtheorem{exam}{\textsc{Example}}

\begin{document}
%
\title{A Survey of Heterogeneous Information Network Analysis}


\author{Chuan~Shi,~\IEEEmembership{Member,~IEEE,}
	Yitong~Li,~\IEEEmembership{}
	Jiawei~Zhang,~\IEEEmembership{}
	Yizhou~Sun,~\IEEEmembership{Member,~IEEE,}
	and~Philip~S.~Yu,~\IEEEmembership{Fellow,~IEEE}
\IEEEcompsocitemizethanks{
\IEEEcompsocthanksitem C. Shi, Y.T. Li are with Beijing University of Posts and Telecommunications, Beijing, China.
E-mail: shichuan@bupt.edu.cn, liyitong@bupt.edu.cn.
\IEEEcompsocthanksitem J.W. Zhang and P.S. Yu are with University of Illinois at Chicago, IL, USA.
E-mail: jwzhanggy@gmail.com, psyu@uic.edu.
\IEEEcompsocthanksitem Y.Z. Sun is with Northeastern University, MA, USA.
E-mail: yzsun@ccs.neu.edu.
}
\thanks{}
}

%
%
%
\markboth{IEEE Transactions on Knowledge and Data Engineering}{Shell \MakeLowercase{\textit{C. Shi et al.}}: A Survey of Heterogeneous Information Network Analysis}




\maketitle

\begin{abstract}
Most real systems consist of a large number of interacting, multi-typed components, while most contemporary researches model them as homogeneous networks, without distinguishing different types of objects and links in the networks. Recently,  more and more researchers begin to consider these interconnected, multi-typed data as heterogeneous information networks, and develop structural analysis approaches by leveraging the rich semantic meaning of structural types of objects and links in the networks. Compared to widely studied homogeneous network, the heterogeneous information network contains richer structure and semantic information, which provides plenty of opportunities as well as a lot of challenges for data mining. In this paper, we provide a survey of heterogeneous information network analysis. We will introduce basic concepts of heterogeneous information network analysis, examine its developments on different data mining tasks, discuss some advanced topics, and point out some future research directions.
\end{abstract}


\section{Introduction}
We know that most real systems usually consist of a large number of interacting, multi-typed components \cite{Han09}, such as human social activities, communication and computer systems, and biological networks. In such systems, the interacting components constitute interconnected networks, which can be called information networks without loss of generality \cite{SH12}. It is clear that information networks are ubiquitous and form a critical component of modern information infrastructure. The information network analysis has gained extremely wide attentions from researchers in many disciplines, such as computer science, social science, physics, and so on. Particularly, the information network analysis has become a hot research topic in data mining and information retrieval fields in the past decades. The basic paradigm is to mine hidden patterns through mining link relations from networked data. The analysis of information network is related to the works in link mining and analysis \cite{GD05,JG98,F02}, social network analysis \cite{W94,OR02}, hypertext and web mining \cite{C02}, network science \cite{L11}, and graph mining \cite{CH00}. 

Most of contemporary information network analyses have a basic assumption: the type of objects or links is unique \cite{SH12}. That is, the networks are homogeneous containing the same type of objects and links, such as the author collaboration network \cite{LLC10} and the friendship network \cite{LCB10}. These homogeneous networks usually are extracted from real interacting systems by simply ignoring the heterogeneity of objects and links or only considering one type of relations among one type of objects. However, most real systems contain multi-typed interacting components and we can model them as heterogeneous information networks \cite{SH12} (called HIN or heterogeneous network for short) with different types of objects and links. For example, in bibliographic database, like DBLP \cite{SH12}, papers are connected together via authors, venues and terms; and in Flickr, photos are linked together via users, groups, tags and comments.

%

Compared to widely-used homogeneous information network, the heterogeneous information network can effectively fuse more information and contain rich semantics in nodes and links, and thus it forms a new development of data mining. More and more researchers have noticed the importance of heterogeneous information network analysis and many novel data mining tasks have been exploited in such networks, such as similarity search \cite{SKYX12,SHYYW11}, clustering \cite{SNHYYY12}, and classification \cite{KYDW12}. Since Y. Sun, J. Han, et al. proposed the concept of heterogeneous information network in 2009 \cite{SHZYCW09}, and the meta path concept subsequently in \cite{SHYYW11}, heterogeneous information network analysis becomes a hot topic rapidly in the fields of data mining, database, and information retrieval, and a lot of papers have appeared in top conferences and journals of these research fields. In addition, some special workshops on heterogeneous information networks began to be held. For example, the workshop on Heterogeneous Information Network Analysis (HINA) has been held for 3 years in conjunction with IJCAI, and the workshop on Mining Data Semantics (MDS) has also been held for several times. 


This paper firstly presents a survey of heterogeneous information network analysis in recent years. Although some articles have introduced the developments of this field \cite{SH12,SH12B,SH13}, they focus on summarizing the works of authors themselves. This paper attempts to clearly introduce basic concepts in heterogeneous network analysis and make a comprehensive investigation on contemporary research developments. Moreover, this paper also discusses some advanced topics and points out several future development directions of this field. 

The remaining part is organized as follows. Section II introduces the basic concepts and examples in this field. Section III presents research developments in  major data mining tasks, and the advanced topics and future works are introduced in Section IV. Finally, Section V concludes this paper.

\section{Basic Concepts and Definitions}
In this section, we introduce some basic concepts in this field, compare the heterogeneous information network with other related concepts, and give some HIN examples. 

\subsection{Basic definitions}
An information network represents an abstraction of the real
world, focusing on the objects and the interactions among these
objects. Formally, we define an information network as follows.

\begin{defn}
	\textbf{Information Network} \cite{SH12,SYH09}. An information network is defined as a directed graph $G = (V,E)$ with an object type mapping function $\varphi : V \rightarrow \mathcal{A}$ and a link type mapping function $\psi : E \rightarrow \mathcal{R}$. Each object $v \in V$ belongs to one particular object type in the object type set $\mathcal{A}$: $\varphi (v) \in \mathcal{A}$, and each link $e \in E$ belongs to a particular relation type in the relation type set  $\mathcal{R}$: $\psi (e) \in \mathcal{R}$. If two links
	belong to the same relation type, the two links share the same
	starting object type as well as the ending object type.
\end{defn}

Different from the traditional network definition, we explicitly distinguish object types and relationship types in information network.

\begin{defn}
	\textbf{Heterogeneous/homogeneous information Network}. The information network is called \textbf{heterogeneous information network} if the types of objects $|\mathcal{A}| > 1$ or the types of relations $|\mathcal{R}| > 1$; otherwise, it is a \textbf{homogeneous information network}.
\end{defn}

\begin{figure}[htbp]
 \centering
 \subfigure[Network instance]{
	\label{fig:DBLP instance}
		\begin{minipage}[t]{0.19\textwidth}
 			\includegraphics[width=3cm]{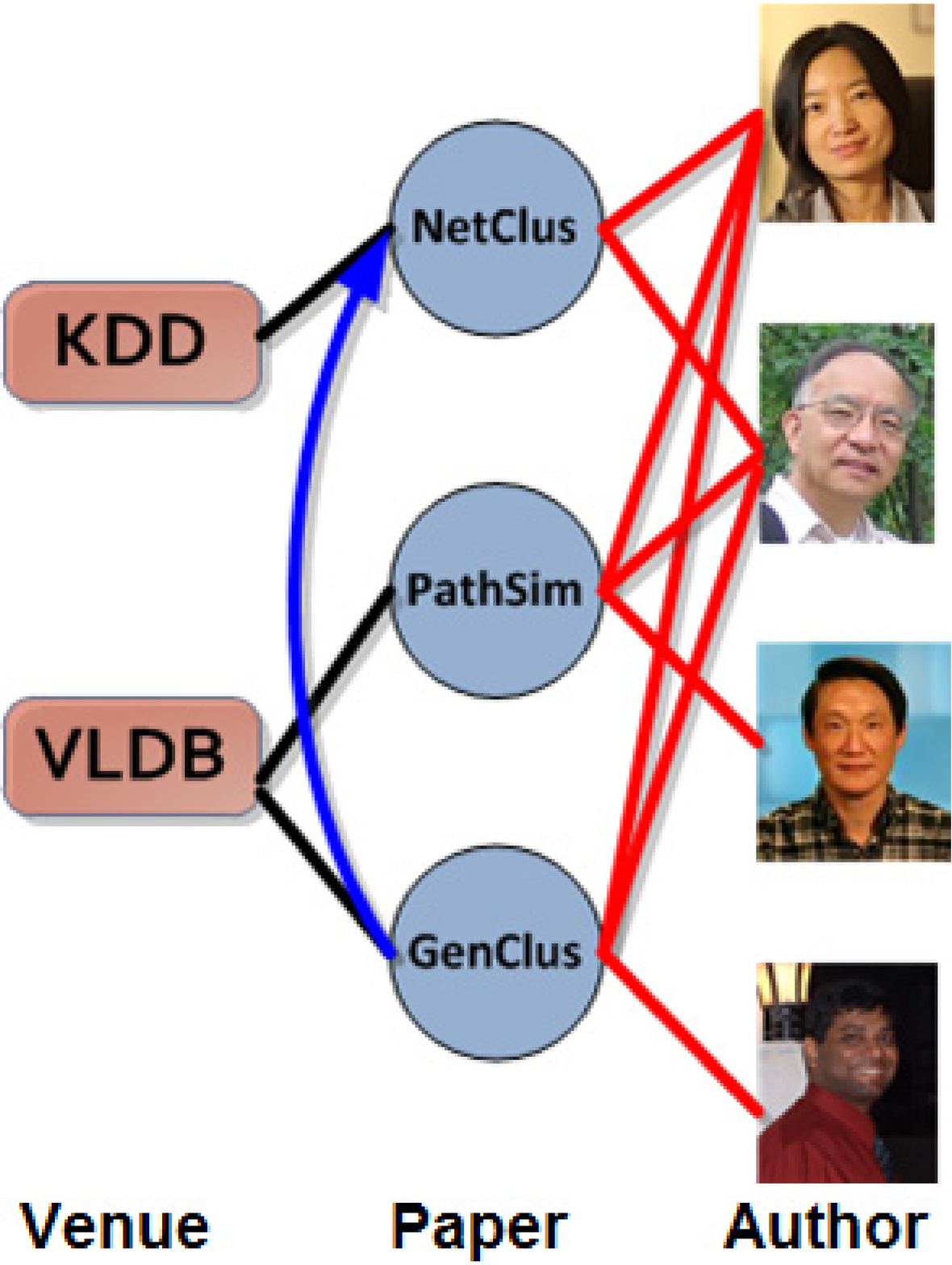}
		\end{minipage}
 }
 \hspace{25pt}
 \subfigure[Network schema]{
	\label{fig:DBLP schema}
		\begin{minipage}[t]{0.31\textwidth}
 			\includegraphics[width=5.4cm]{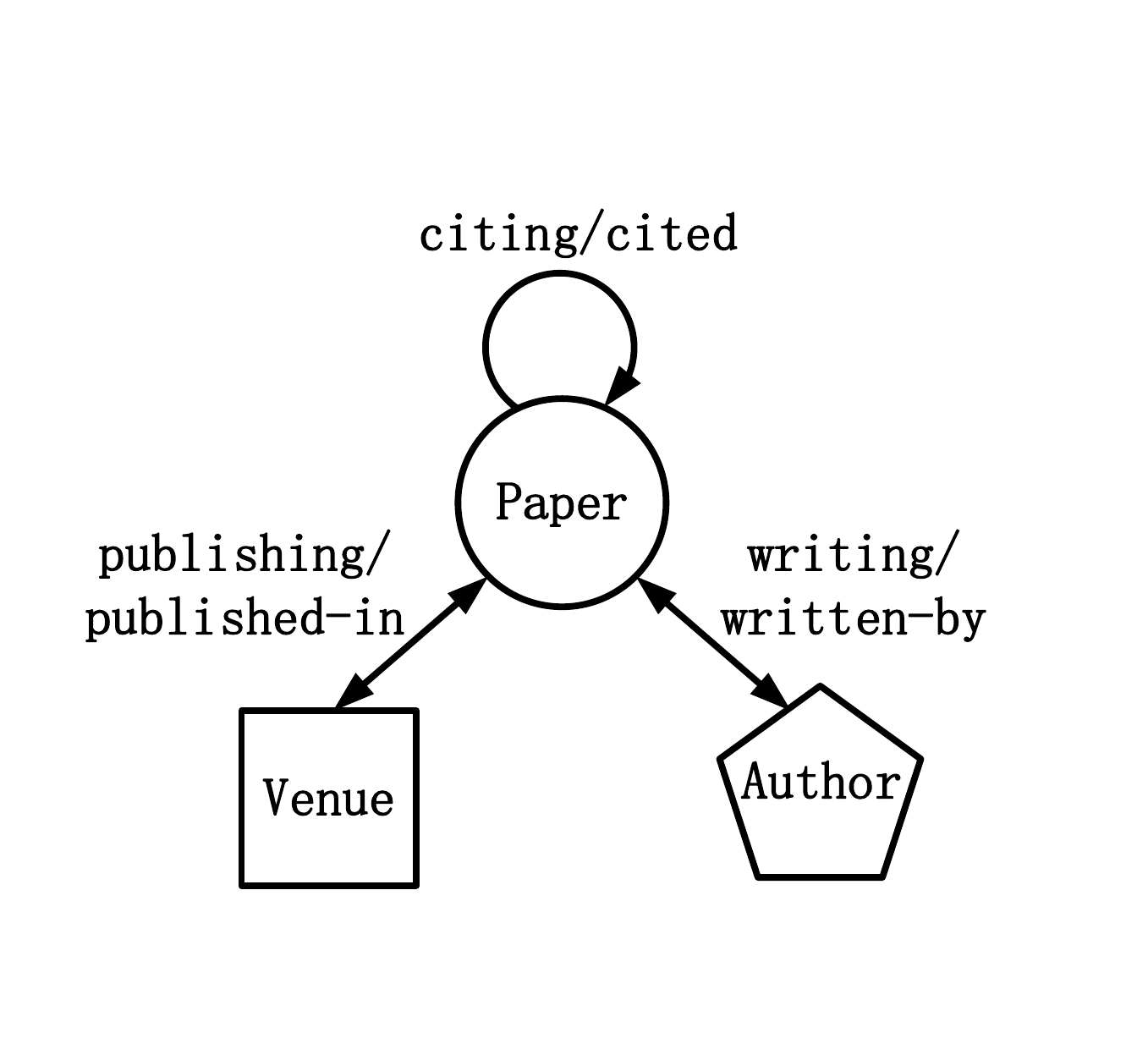}
		\end{minipage}
}
 \caption{An example of heterogeneous information network on bibliographic data \cite{SH12}.}
 \label{fig:bib}
\end{figure}

%
%

\begin{exam}
	Fig. \ref{fig:bib} shows a HIN example on bibliographic data \cite{SH12}. A bibliographic information network, such as the bibliographic network involving computer science researchers derived from DBLP \footnote{http://dblp.uni-trier.de/}, is a typical heterogeneous network containing three types of information entities: papers, venues, and authors. For each paper, it has links to a set of authors, and a venue, and these links belong to a set of link types. 
%
%
%
\end{exam}

In order to better understand the object types and link
types in a complex heterogeneous information network, it is
necessary to provide the meta level (i.e., schema-level) description of the network. Therefore, the concept of
network schema is proposed to describe the meta structure of a network.

\begin{defn}
	\textbf{Network schema} \cite{SH12,SYH09}. The network
	schema, denoted as $T_G =(\mathcal{A},\mathcal{R})$, is a meta template for
	an information network $G = (V, E)$ with the object type
	mapping $\varphi : V \rightarrow \mathcal{A}$ and the link type mapping $\psi : E \rightarrow \mathcal{R}$, which
	is a directed graph defined over object types $\mathcal{A}$, with edges
	as relations from $\mathcal{R}$.
\end{defn}

The network schema of a heterogeneous information network specifies type constraints on the sets of objects and relationships among the objects. These constraints make
a heterogeneous information network semi-structured, guiding the semantics explorations of the network. An information network following a network schema is called
a \textbf{network instance} of the network schema. For a
link type $R$ connecting object type $S$ to object type $T$, i.e., $S\overset{R}{\longrightarrow}T$, $S$ and $T$ are the \textbf{source
object type} and \textbf{target object type} of link type $R$, which can be denoted as
$R.S$ and $R.T$, respectively. The inverse relation $\emph{R}^{-1}$
holds naturally for $T\overset{\emph{R}^{-1}}{\longrightarrow}S$.
Generally, $\emph{R}$ is not equal to $\emph{R}^{-1}$, unless
$\emph{R}$ is symmetric.

\begin{exam}
	As described above, Fig. \ref{fig:bib}(a) demonstrates the real objects and their connections on bibliographic data. Fig. \ref{fig:bib}(b) illustrates its network schema which describes the object types and their relations in the HIN. Moreover, Fig. \ref{fig:bib}(a) is a network instance of the network schema Fig. \ref{fig:bib}(b).  In this example, it contains objects from three types of objects: papers ($P$), authors ($A$), and venues ($V$). There are links connecting different types of objects. The link types are defined by the relations between two object types. For example, links existing between authors and papers denote the writing or written-by relations, while those between venues and papers denote the publishing or published-in relations.
\end{exam}

Different from homogeneous networks, two objects in a heterogeneous network can be connected via different paths and these paths have different physical meanings. These paths can be categorized as meta paths as follows.

\begin{figure}[htbp]
 \centering
\subfigure[APA]{
\label{fig:apa}
\begin{minipage}[t]{0.23\textwidth}
  \includegraphics[width=3.2cm]{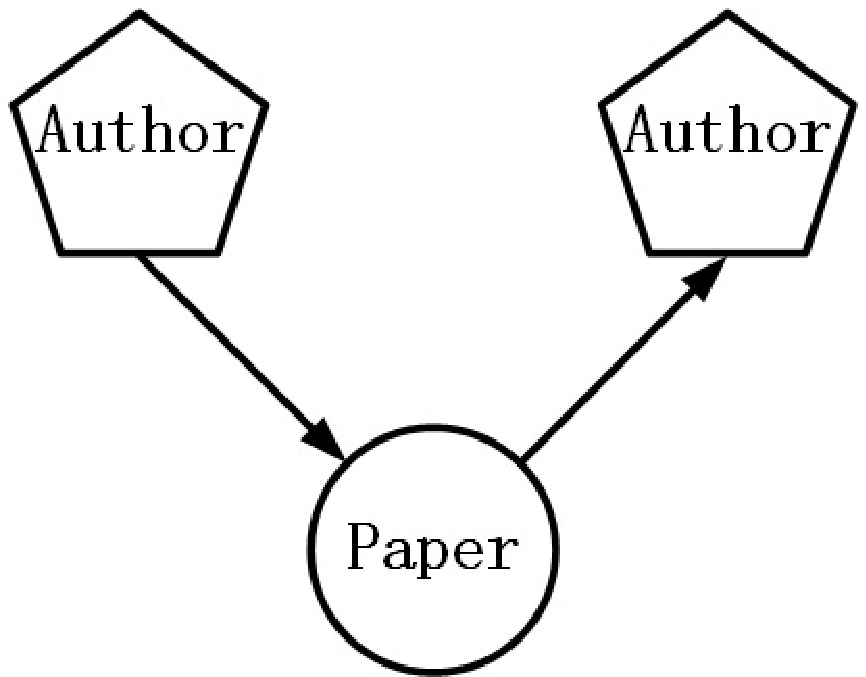}
\end{minipage}
}
\subfigure[APVPA]{
\label{fig:apvpa}
	\begin{minipage}[t]{0.23\textwidth}
		\includegraphics[width=3.2cm]{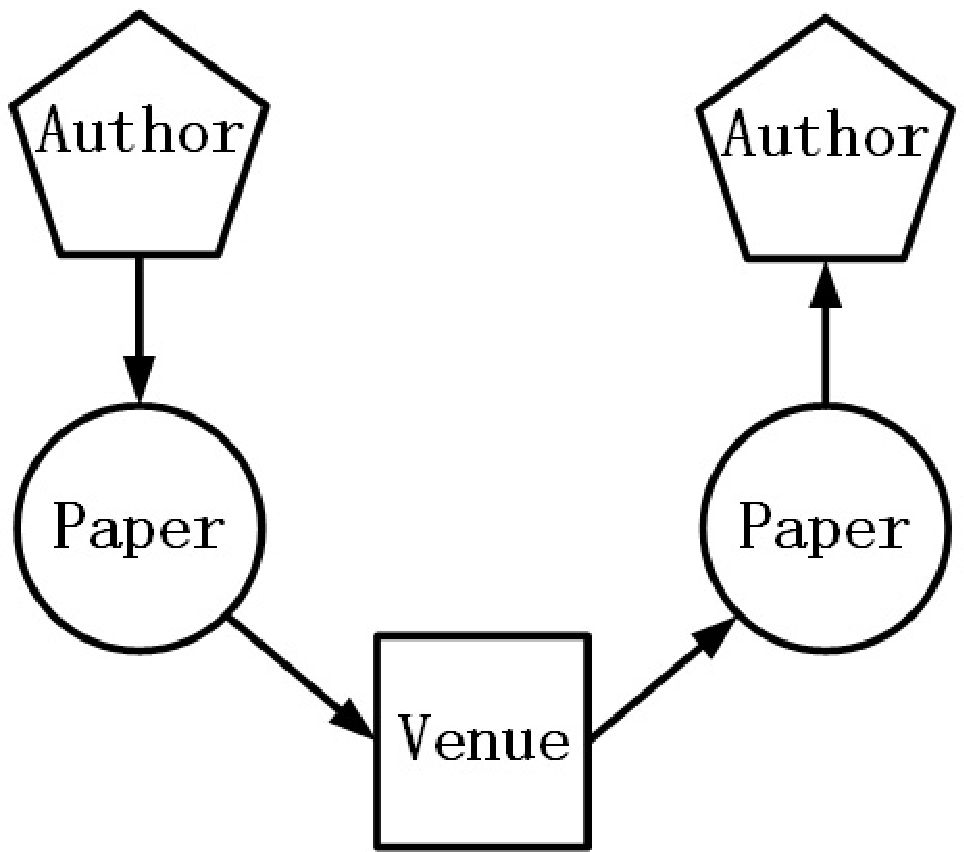}
	\end{minipage}
}
\subfigure[APV]{
\label{fig:apv}
	\begin{minipage}[t]{0.23\textwidth}
		\includegraphics[width=3.2cm]{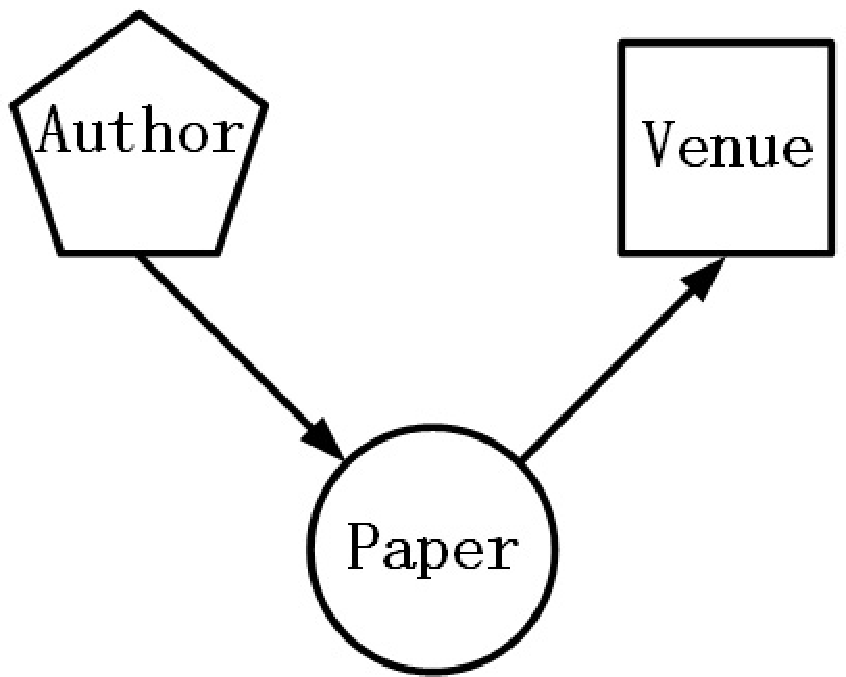}
	\end{minipage}
}
\caption{Examples of meta paths in heterogeneous information network on bibliographic data.}
\label{fig:pathschema}
\end{figure}

\begin{defn}
	\textbf{Meta path} \cite{SHYYW11}. A meta path $\mathcal{P}$ is a path defined on a schema $S=(\mathcal{A},\mathcal{R})$, and is denoted in the form of $A_1 \xrightarrow{R_1} A_2 \xrightarrow{R_2} \ldots \xrightarrow{R_l} A_{l+1}$, which defines a composite relation $R = R_1 \circ R_2 \circ \cdots \circ R_l$ between objects $A_1, A_2, \cdots, A_{l+1}$, where $\circ$ denotes the composition operator on relations.
\end{defn}

For simplicity, we can also use object types to denote the meta
path if there are no multiple relation types between the same pair of
object types: $\mathcal{P}=(A_1A_2\cdots A_{l+1})$. For example, in Fig. 
\ref{fig:bib}(a), the relation, authors publishing papers in
conferences, can be described using the length-2 meta path
$A\overset{writting}{\longrightarrow}P\overset{written by}{\longrightarrow}A$,
or $APA$ for short. We say a concrete path $p=(a_1a_2\cdots a_{l+1})$
between objects $a_1$ and $a_{l+1}$ in network $G$ is a \textbf{path
instance} of the relevance path $\mathcal{P}$, if for each $a_i$,
$\phi(a_i)=A_i$ and each link $e_i=\langle a_i, a_{i+1}\rangle$
belongs to the relation $R_i$ in $\mathcal{P}$. It can be denoted as
$p\in \mathcal{P}$. A meta path $\mathcal{P}$ is a \textbf{symmetric path}, if the relation $R$
defined by it is symmetric (i.e., $\mathcal{P}$ is equal to
$\mathcal{P}^{-1}$), such as $APA$ and $APVPA$. Two meta paths
$\mathcal{P}_1=(A_1A_2\cdots A_l)$ and $\mathcal{P}_2=(B_1B_2\cdots
B_k)$ are \textbf{concatenable} if and only if $A_l$ is equal to $B_1$, and the
concatenated path is written as $\mathcal{P}=(\mathcal{P}_1\mathcal{P}_2)$, which equals to
$(A_1A_2\cdots A_lB_2\dots B_k)$. A simple concatenable example is
that $AP$ and $PA$ can be concatenated to the path $APA$.

\begin{table}[htbp]
  \label{table:path}
  \caption{Meta path examples and their physical meanings on bibliographic data.}
\begin{center}
\resizebox{.95\textwidth}{!}
{
    \begin{tabular}{|c|c|c|}
    \hline
      Path instance & Meta path & Physical meaning \\
    \hline
      Sun-NetClus-Han & \multirow{2}[2]{*}{Author-Paper-Author (\emph{APA})} & \multirow{2}[2]{*}{Authors collaborate on the same paper} \\
      Sun-PathSim-Yu &{}&{}\\
    \hline
      Sun-PathSim-VLDB-PathSim-Han & \multirow{2}[2]{*}{Author-Paper-Venue-Paper-Author (\emph{APVPA})} & \multirow{2}[2]{*}{Authors publish papers on the same venue} \\
      Sun-PathSim-VLDB-GenClus-Aggarwal &{}&{}\\
    \hline
      Sun-NetClus-KDD & \multirow{2}[2]{*}{Author-Paper-Venue (\emph{APV})} & \multirow{2}[2]{*}{Authors publish papers at a venue} \\
      Sun-PathSim-VLDB &{}&{}\\
    \hline
    \end{tabular}%
}
\end{center}
\end{table}%

\begin{exam}
	As examples shown in Fig. \ref{fig:pathschema}, authors can be connected via meta paths ``Author-Paper-Author'' ($APA$) path, ``Author-Paper-Venue-Paper-Author'' ($APVPA$) path, and so on. Moreover, TABLE I shows path instances and semantics of these meta paths.  It is obvious that semantics underneath these paths are different. The $APA$ path means authors collaborating on the same papers (i.e., co-author relation), while $APVPA$ path means authors publishing papers on the same venue. The meta paths can also connect different types of objects. For example, the authors and venues can be connected with the $APV$ path, which means authors publishing papers on venues. 
\end{exam}

The rich semantics of meta path is an important characteristic of HIN. Based on different meta paths, objects have different connection relations with diverse path semantics, which may have an effect on many data mining tasks. For example, the similarity scores among authors evaluated based on different meta paths are different \cite{SHYYW11}. Under the \emph{APA} path, the authors co-publishing papers will be more similar, while the authors publishing papers on the same venues will be more similar under the \emph{APVPA} path. Another example is the importance evaluation of objects \cite{LSPC14}. The importance of authors under \emph{APA} path has a bias on the authors who write many papers having many authors, while the importance of authors under \emph{APVPA} path emphasizes the authors who publish many papers on those productive conferences. As a unique characteristic and effective semantic capturing tool, meta path has been widely used in many data mining tasks in HIN, such as similarity measure \cite{SKYX12,SHYYW11}, clustering \cite{SNHYYY12}, and classification \cite{KYDW12}. 

\subsection{Comparisons with related concepts}

With the boom of social network analysis, all kinds of networked data have emerged, and numbers of concepts to model networked data have been proposed. Here we compare heterogeneous network concept with these related concepts.   

\textbf{Heterogeneous network vs homogeneous network}. Heterogeneous networks include different types of nodes or links, while homogeneous networks only have one type of objects and links. Homogeneous networks can be considered as a special case of heterogeneous networks. Moreover, a heterogeneous network can be converted into a homogeneous network through network projection or ignoring object heterogeneity, while it will make significant information loss. Traditional link mining \cite{TGHL13,KSJ09,LK03} is usually based on homogeneous network, and many analysis techniques on homogeneous network cannot be directly applied to heterogeneous network. 

\textbf{Heterogeneous network vs multi-relational network \cite{YCSH12}}. Different from heterogeneous network, multi-relational network has only one type of objects, but more than one kind of relationship between objects. So multi-relational network can be seen as a special case of heterogeneous network. 

\textbf{Heterogeneous network vs multi-dimensional/mode network \cite{TLZN08}}. Tang et al. \cite{TLZN08} proposed the multi-dimensional/mode network concept, which has the same meaning with multi-relational network. That is, the network has only one type of objects and more than one kind of relationship between objects. So multi-dimensional/mode network is also a special case of heterogeneous network. 

\textbf{Heterogeneous network vs composite network \cite{ZFWXL12,ZFZY13}.} Qiang Yang et al. proposed the composite network concept \cite{ZFWXL12,ZFZY13}, where users in networks have various relationships, exhibit different behaviors in each individual network or subnetwork, and share some common latent interests across networks at the same time. So composite network is in fact a multi-relational network, a special case of heterogeneous network.

\textbf{Heterogeneous network vs complex network.} A complex network is a network with non-trivial topological features and patterns of connection between its elements that are neither purely regular nor purely random \cite{KW08}. Such non-trivial topological features include a heavy tail in the degree distribution, a high clustering coefficient, community structure, and hierarchical structure. The studies of complex networks have brought together researchers from many areas including mathematics, physics, biology, computer science, sociology, and others. The studies show that many real networks are complex networks, such as social networks, information networks, technological networks, biological networks, and so on \cite{N03}. So we can say that many real heterogeneous networks are complex networks. However, the studies on complex networks usually focus on the structures, functions, and features of networks.

\subsection{Example datasets of heterogeneous information network}
Intuitively, most real systems include multi-typed interacting objects. For example, a social media website (e.g., Facebook) contains a set of object types, such as users, posts, and tags, and a health care system contains doctors, patients, diseases, and devices. Generally speaking, these interacting systems can all be modeled as heterogeneous information networks. Concretely, this kind of networks can be constructed from the following three types of data.

1) Structured data. Structured data stored in database table is organized with entity-relation model. The different-typed entities and their relations naturally construct information networks. For example, the bibliographic data (see the above example) is widely used as heterogeneous information network.

2) Semi-structured data. Semi-structured data is usually stored with XML format. The attributes in XML can be considered as object types, and the object instances can be determined by analyzing the contents of attributes. The connections among attributes construct object relations.

3) Non-structured data. For non-structured data, heterogeneous information networks can also be constructed by objects and relationship extraction. For example, for text data, entity recognition and relation extraction can form the objects and links of HIN.

\begin{figure}[htbp]
	\centering
	\subfigure[Multi-relation]{
	\label{fig:xiaonei}
		\begin{minipage}[t]{0.17\textwidth}
 			\includegraphics[width=2.7cm]{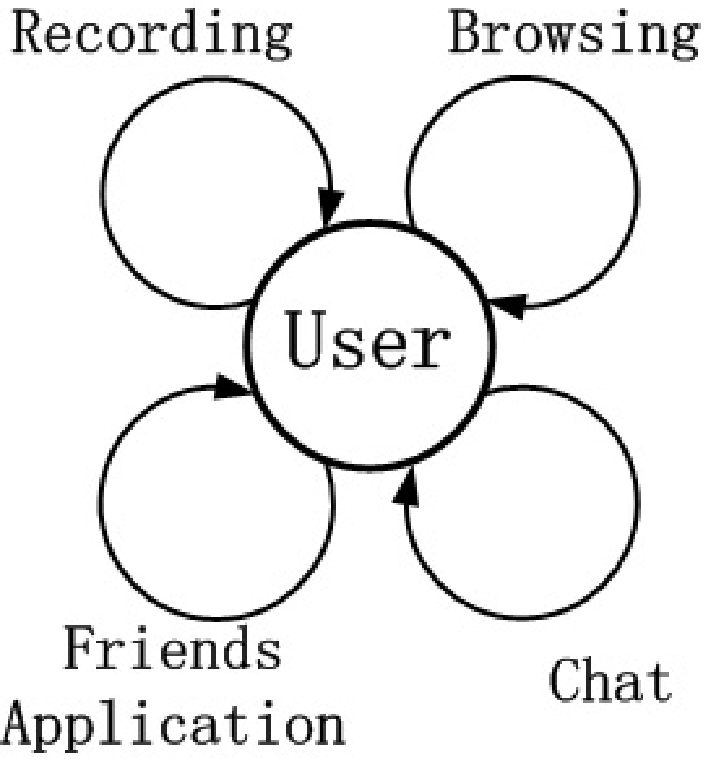}
		\end{minipage}
	}
	\subfigure[Bipartite]{
	\label{fig:ng20}
		\begin{minipage}[t]{0.08\textwidth}
  			\includegraphics[width=1.1cm]{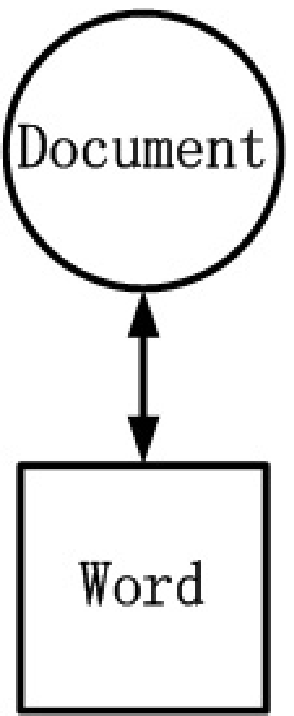}
		\end{minipage}
	}
	\subfigure[Star-schema]{
	\label{fig:dblp}
		\begin{minipage}[t]{0.18\textwidth}
  			\includegraphics[width=3cm]{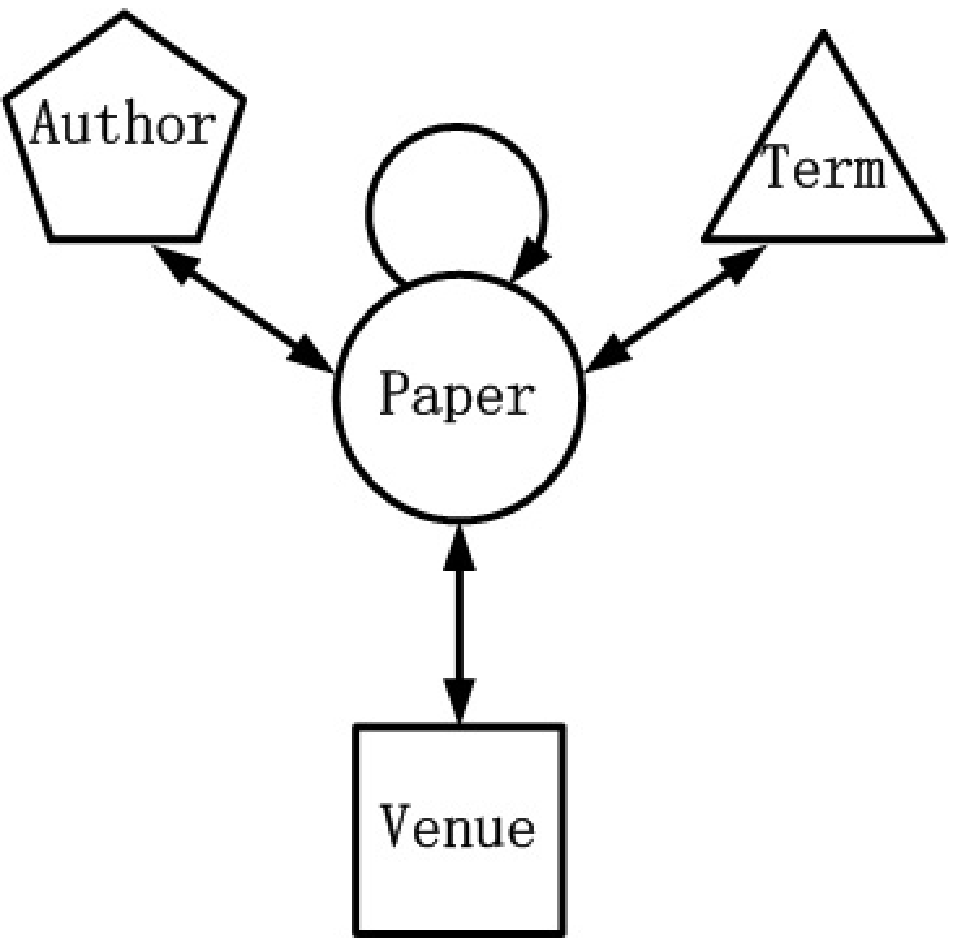}
		\end{minipage}
	}	
	\subfigure[Multiple-hub]{
	\label{fig:slap}
		\begin{minipage}[t]{0.23\textwidth}
  			\includegraphics[width=3.9cm]{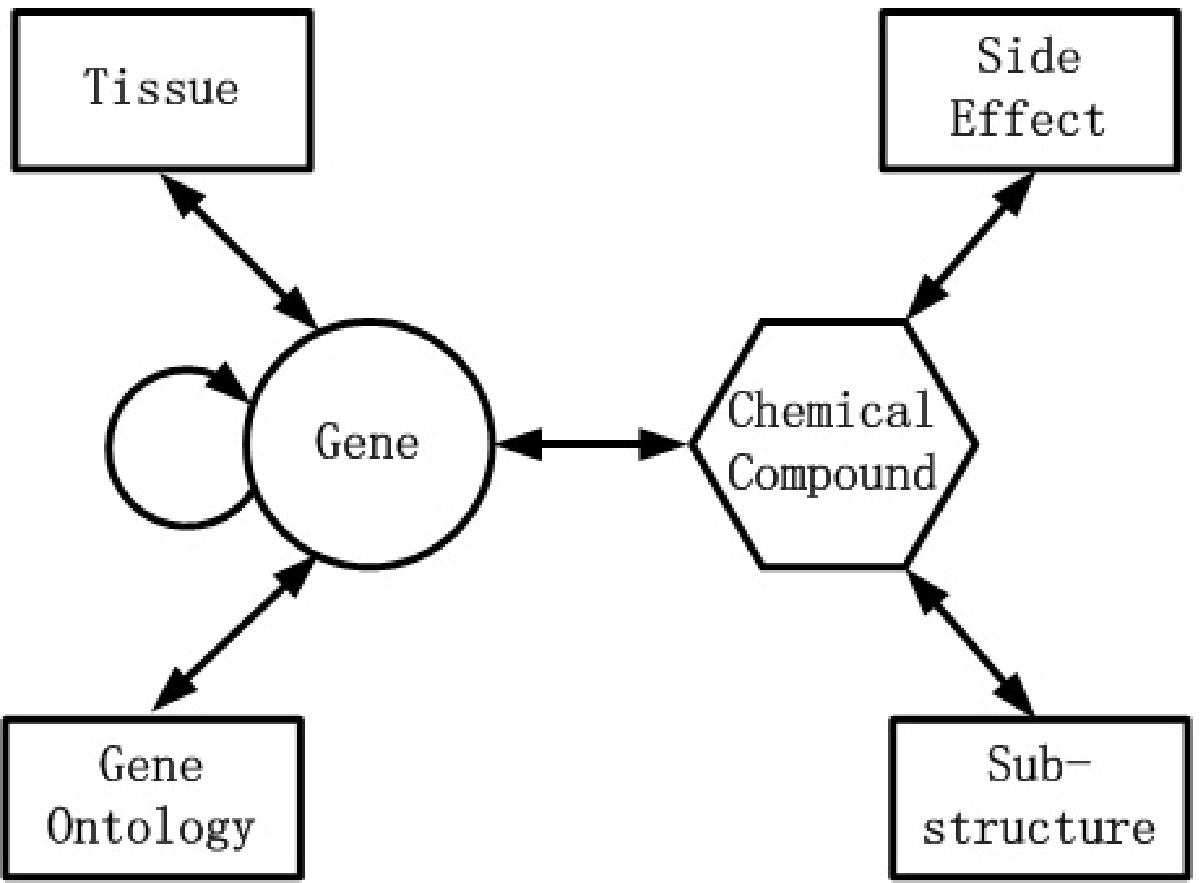}
		\end{minipage}
	}
	\caption{Network schema of heterogeneous information networks.}
\end{figure}

Although heterogeneous information networks are ubiquitous, there are not many standard datasets for study. Here we summarize some widely used heterogeneous networks in literals.

\textbf{Multi-relational network with single-typed object}. Traditional multi-relational network is a kind of HIN, where there is one type of object and several types of relations among objects. This kind of networks widely exist in social websites, like Facebook and Xiaonei \cite{ZFZY13}. Fig. \ref{fig:xiaonei} shows the network schema of such a network \cite{ZFZY13}, where users can be extensively connected with each other through connections, such as recording, browsing, chatting, and sending friends applications. 

\textbf{Bipartite network}. As a typical HIN, bipartite network is widely used to construct interactions among two types of objects, such as user-item \cite{JL13}, and document-word \cite{LZY05}. Fig. \ref{fig:ng20} shows the schema of a bipartite network connecting documents and words \cite{LZY05}. As an extension of bipartite graphs, $k$-partite graphs \cite{LWZY06} contain multiple types of objects where links exist among adjacent object types. 

\textbf{Star-schema network}. Star-schema network is the most popular HIN in this field. In database table, a target object and its attribute objects naturally construct a HIN, where the target object, as the hub node, connects different attribute objects. As an example shown in Fig. \ref{fig:dblp}, a bibliographic information network is a typical star-schema heterogeneous network \cite{SHYYW11,SKYX12}, containing different objects (e.g., paper, venue, author, and term) and links among them. Many other datasets can also be represented as star-schema networks, such as the movie data \cite{SZKY12, YRSSKGNH13} from the Internet Movie Database \footnote{www.imdb.com/} (IMDB) and the patent data \cite{ZZBTCYH14} from US patents data \footnote{http://www.uspto.gov/patents/}. 

\textbf{Multiple-hub network}. Beyond star schema, some networks have more complex structures, which involve multiple hub objects. This kind of networks widely exist in bioinformatics data \cite{KCY13,WSYW13}. A bioinformatics example, shown in Fig. \ref{fig:slap}, includes two hubs: Gene and Chemical compound. Another example can be found in the Douban dataset \footnote{http://www.douban.com/} \cite{SZLYYW15}.



\begin{figure}[htbp]
 \centering
\subfigure[Multiple HINs \cite{KZY13}]{
\label{fig:apa}
\begin{minipage}[t]{0.34\textwidth}
 \includegraphics[width=6cm]{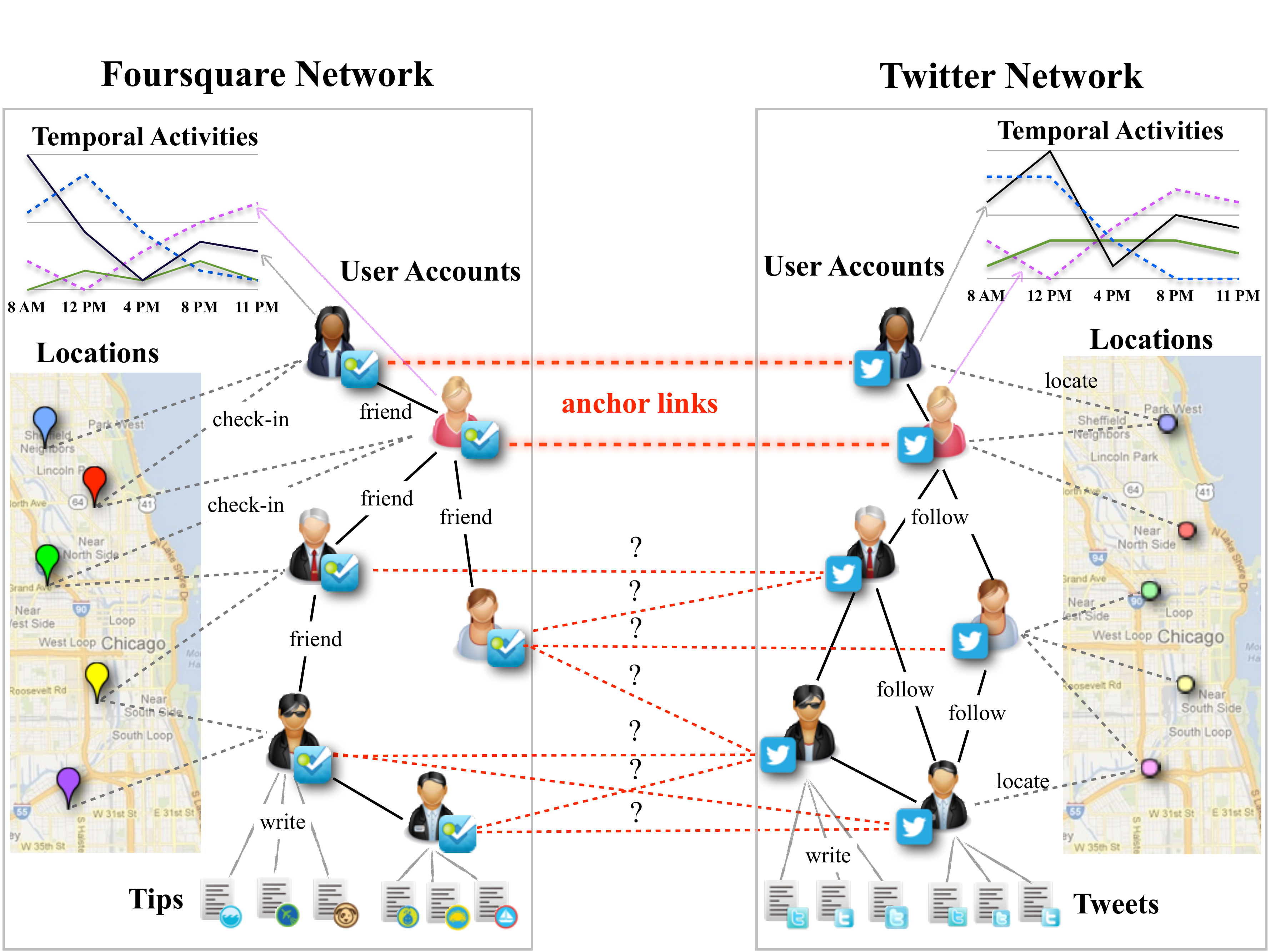}
\end{minipage}
}
\hspace{50pt}
\subfigure[Schema-rich HIN \cite{ZOCSHZ14}]{
\label{fig:apvpa}
	\begin{minipage}[t]{0.4\textwidth}
		 \includegraphics[width=7.2cm]{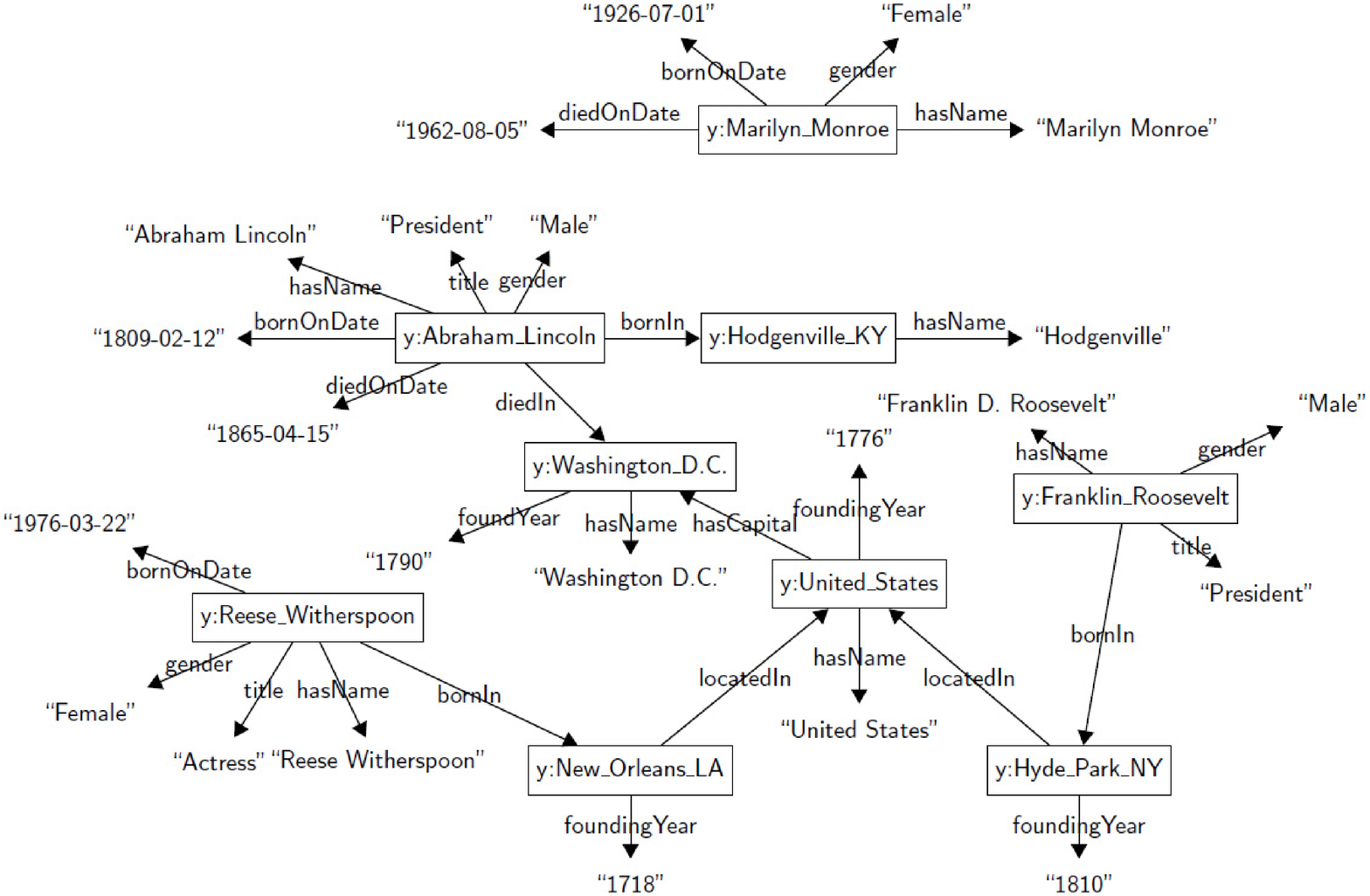}
	\end{minipage}
}
\caption{Two examples of complex heterogeneous information network.}
\label{fig:complexHIN}
\end{figure}

Besides these widely used networks, many real systems can also be constructed as more complex heterogeneous networks. In some real applications, users may  exist in multiple social networks, and each social network can be modeled as an HIN. Fig. \ref{fig:complexHIN}(a) shows an example of two heterogeneous social networks (Twitter and Foursquare) \cite{KZY13}. In each network, users are connected with each other through social links, and they are also connected with a set of locations, timestamps and text contents through online activities. Moreover, some users have two accounts in two social networks separately, and they serve as anchor nodes to connect two networks.  More generally, some interaction systems are too complex to be modeled as an HIN with a simple network schema. Knowledge graph \cite{S12} is such an example. We know that knowledge graph is based on Resource Description Framework (RDF) data, which complies with an $<Subject, Property, Object>$ model. Here ``Subject" and ``Object" can be considered as objects, and ``Property" can be considered as the relation between ``Subject" and ``Object". And thus a knowledge graph can be considered as a heterogeneous network, an example is shown in Fig. \ref{fig:complexHIN}(b). In such a semantic knowledge base, like Yago \cite{SKW07}, there are more than 10 million entities (or nodes) of different types, and more than 120 million links among these entities. In such a schema-rich network, it is impossible to depict such network with a simple network schema. 

%

%

%

\subsection{Why Heterogeneous Information Network Analysis}
In the past decades, link analysis has been extensively explored \cite{GD05}. So many methods have been developed for information network analysis and numerous data mining tasks have been explored in homogeneous networks, such as ranking, clustering, link prediction, and influence analysis. However, due to some unique characteristics (e.g., fusion of more information and rich semantics) of HIN, most methods in homogeneous networks cannot be directly applied in heterogeneous networks, and it is potential to discover more interesting patterns in this kind of networks. 

\textbf{It is a new development of data mining.} Early data mining problems focused on analyzing feature vectors of objects. In the late 1990s, with the advent of WWW, more and more data mining researches turned to study links among objects. It is one of the main research directions to mine hidden patterns from feature and link information of objects. In these researches, homogeneous networks are usually constructed from interconnected objects. In recent years, abundant social media emerge, and many different types of objects are interconnected. It is hard to model these interacted objects as homogeneous networks, while it is natural to model different types of objects and relations among them as heterogeneous networks. Particularly, with the rapid increment of user-generated content online, big data analysis is an emergent yet important task to be studied. Variety is one significant characteristic of big data \cite{WZWD14}. As a semi-structured representation, heterogeneous information network can be an effective tool to deal with complex big data.  

\textbf{It is an effective tool to fuse more information.} Compared to homogeneous network, heterogeneous network is natural to fuse more objects and their interactions. In addition, traditional homogeneous networks are usually constructed from single data source, while heterogeneous network can fuse information across multiple data sources. For example, customers use many services provided by Google, such as Google search, G-mail, maps, Google+, etc. So we can fuse these information with a heterogenous information network, in which customers interact with many different types of objects, such as  key words, mails, locations, followers, etc. Broadly speaking, heterogeneous information network can also fuse information cross multiple social network platforms.  We know that there are many social network platforms with different objectives, such as Facebook, Twitter, Weixin, and Weibo. Moreover, users often participate in multiple social networks. Since each social network only captures a partial or biased view of a user, we can fuse information across multiple social network platforms with multiple heterogeneous information networks, where each heterogeneous network represents information from one social network with some anchor nodes connecting these networks.    

\textbf{It contains rich semantics.} In heterogeneous networks, different-typed objects and links coexist and they carry different semantic meanings. As a bibliographic example shown in Fig. \ref{fig:bib}, it includes author, paper, and venue object types. The relation type ``Author-Paper'' means author writing paper, while the relation type ``Paper-Venue'' means paper published in venue. Considering the semantic information will lead to more subtle knowledge discovery. For example, in DBLP bibliographic data \cite{SHYYW11}, if you find the most similar authors to ``Christos Faloutsos'', you will get his students, like Spiros Papadimitriou and Jimeng Sun, under the $APA$ path; while the results are reputable researchers, like Jiawei Han and Rakesh Agrawal, under the $APVPA$ path. How to mine interesting patterns with the semantic information is a unique issue in heterogeneous network. 

\section{Research Developments}

\begin{figure}[htbp]
	\centering
	\includegraphics[width=10cm]{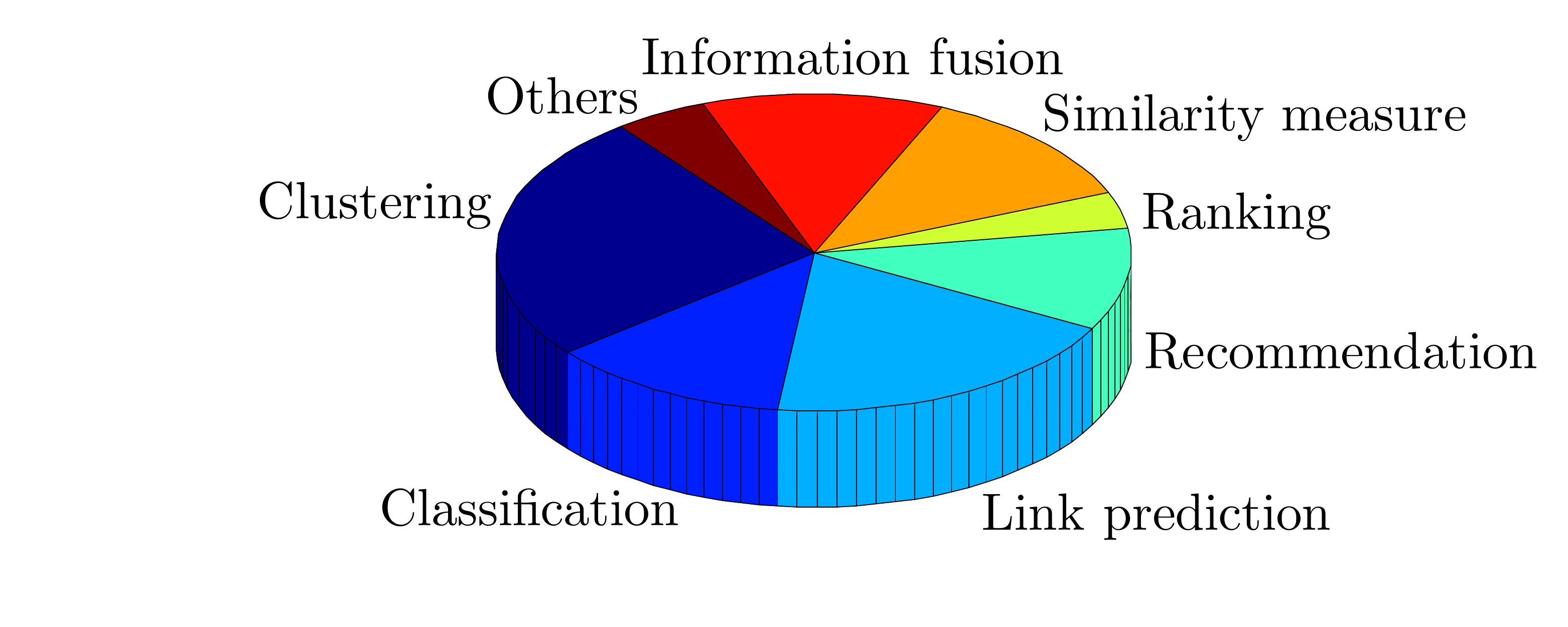}
	\caption{Paper distribution of heterogeneous information network analysis on different data mining tasks.}
	\label{fig:field1}
\end{figure}

Heterogeneous information network provides a new paradigm to manage networked data. Meanwhile, it also introduces new challenges for many data mining tasks. We have analyzed more than 100 papers in this field, and divided them into 7 categories according to their data mining tasks. The proportion of papers belonging to each category is shown in Fig. \ref{fig:field1}. In this section, we will summarize the developments about these 7 main data mining tasks.  

\subsection{Similarity Measure}

Similarity measure is to evaluate the similarity of objects. It is the basis of many data mining tasks, such as web search, clustering, and product recommendation. Similarity measure has been well
studied for a long time. These studies can be roughly categorized into two types: feature based approaches and link based approaches. The feature based approaches measure the similarity of objects based
on their feature values, such as cosine similarity, Jaccard
coefficient, and Euclidean distance. The link based approaches measure the similarity of objects based on
their link structures in a graph. For example, Personalized PageRank \cite{JW03} evaluates the
probability starting from a source object to a target object by randomly walking with restart, and SimRank \cite{JW02} evaluates the similarity of two objects by their neighbors' similarities. 



\begin{table}
 \caption{Top-5 most similar authors to ``Christos Faloutsos" under different meta paths on DBLP dataset.}
 \label{tabel:pathsim}
 \centering
        \begin{tabular}{|c|c|c|}
         \hline
         \multirow{2}{*}{Rank} & 
         \multicolumn{2}{c|}{Authors}\\
		 \cline{2-3}
			& \emph{APA} & \emph{APCPA}\\
         \hline
         1 & Christos Faloutsos & Christos Faloutsos\\
         2 & Spiros Papadimitriou & Jiawei Han\\
         3 & Jimeng Sun & Rakesh Agrawal\\
         4 & Jia-Yu Pan & Jian Pei\\
         5 & Agma J. M. Traina & Charu C. Aggarwal\\
         \hline
        \end{tabular}
\end{table}

Recently, many researchers begin to consider similarity measure on heterogeneous information networks. Different from similarity measure on homogeneous networks, similarity measure on HIN not only considers structure similarity of two objects but also takes the meta path connecting these two objects into account. As we know, there are different meta paths connecting two objects, and these meta paths contain different semantic meanings, which may lead to different similarities. And thus the similarity measure on HIN is meta path constraint. For example, based on the bibliographic network in Fig. \ref{fig:bib}, TABLE \ref{tabel:pathsim} shows the most similar authors to Christos Faloutsos, a well-known expert in data mining, under different meta paths \cite{SHYYW11}. Based on \emph{APA} path, the most similar authors to Christos are his students (e.g., Spiros Papadimitriou and Jimeng Sun), while the most similar authors are reputable researchers in the same field with Christos under the \emph{APVPA} path (e.g., Jiawei Han and  Rakesh Agrawal). 

Considering semantics in meta paths constituted by different-typed objects, Sun et al. \cite{SHYYW11} first propose the path based similarity measure PathSim to evaluate the similarity of same-typed objects based on symmetric paths. Following their work, some researchers \cite{HBZ14,HYM14} extend PathSim by incorporating richer information, such as transitive similarity, temporal dynamics, and supportive attributes. A path-based similarity join method \cite{XZY15} is proposed to return the top \emph{k} similar pairs of objects based on user specified join paths. In information retrieval community, Lao and Cohen \cite{LC10,LC10FAST} propose a Path Constrained Random Walk (PCRW) model to measure the entity proximity in a labeled directed graph constructed by the rich metadata of scientific literature. 

In order to evaluate the relevance of different-typed objects, Shi et al. \cite{SKYX12,SKHYW14} propose HeteSim to measure the relevance of any object pair under arbitrary meta path. As an adaption of HeteSim, LSH-HeteSim \cite{LSXZ14} is proposed to mine the drug-target interaction in heterogeneous biological networks where drugs and targets are connected with complicated semantic paths. In order to overcome the shortcoming of HeteSim in high computation and memory demand, Meng et al. \cite{MSLZW14} propose the AvgSim measure that evaluates similarity score through two random walk processes along the given meta path and the reverse meta path, respectively. In addition, some methods \cite{BHPL14,ZZPYXWWH15} combine meta path based relevance search with user preference. 

More works begin to integrate the network structure and other information to measure similarity of objects in HIN. Combining the influence and similarity information, Wang et al. \cite{WHY12} simultaneously measure social influence and object similarity in a heterogeneous network to produce more meaningful similarity scores. Wang et al. \cite{WRFZHB11} propose a model to learn relevance through analyzing the context of heterogeneous networks for online targeting. Yu et al. \cite{YSNMH12} predict the semantic meaning based on a user's query in the meta-path-based feature space and learn a ranking model to answer the similarity query. Recently, Zhang et al. \cite{ZHHW15} propose a similarity measure to compute similarity between centers in an x-star network according to the attribute similarities and the connections among centers. 

\subsection{Clustering}

Clustering analysis is the process of partitioning a set of data objects (or observations) into  a set of clusters, such that objects in a cluster are similar to one another, yet dissimilar to objects in other clusters. Conventional clustering is based on the features of objects, such as k-means and so on \cite{J10}. Recently, clustering based on networked data (e.g., community detection) has been studied a lot. This kind of methods model the data as a homogeneous network, and use the given measure (e.g., normalized cuts \cite{SM00}, and modularity \cite{NG04}) to divide the network into a series of subgraphs. Many algorithms have been proposed to solve this NP-hard problem, such as spectral method \cite{L07}, greedy method \cite{WT07} and sampling technique \cite{SGMA07}. Some researches also simultaneously consider objects' link structure and attribute information to increase the accuracy of clustering \cite{ZCY09,YJCZ09}. 


\begin{figure}[htbp]
	\centering
	\includegraphics[width=10cm]{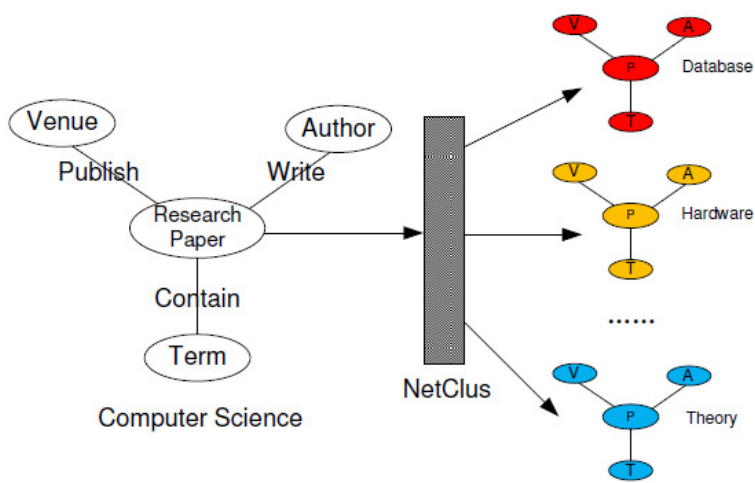}
	\caption{Clustering on a bibliographic heterogeneous network \cite{SYH09}. }
	\label{fig:netclus}
\end{figure}

Recently, clustering of heterogeneous networks attracts much attention. Compared with homogeneous networks, heterogeneous networks integrate multi-typed objects, which generates new challenges for clustering tasks. On the one hand, multiple types of objects co-existing in a network lead to new clustering paradigms. For example, a cluster may include different types of objects sharing the same topic \cite{SYH09}. Fig. \ref{fig:netclus} shows the clustering process on a bibliographic heterogeneous network, which splits the original network into several layers (a set of sub-network clusters). For example, a cluster of the database area consists of a set of database authors, conferences, terms, and papers. In this way, clustering in HIN preserves richer information, but it also faces more challenges. On the other hand, abundant information contained in HIN makes it more convenient to integrate additional information or other learning tasks for clustering. In this section, we will review these works according to the types of integrated information or tasks.

The attribute information is widely integrated into clustering analysis on HIN. Aggarwal et al. \cite{AXY11} use the local succinctness property to create balanced communities across a heterogeneous network. Considering the incompleteness of objects' attributes and different types of links in heterogeneous information networks, Sun et al. \cite{SAH12} propose a model-based clustering algorithm to integrate the incomplete attribute information and the network structure information. Qi et al. \cite{QAH12} propose a clustering algorithm based on heterogeneous random fields to model the structure and content of social media networks with outlier links. Cruz et al. \cite{CBP13} integrate structural dimension and compositional dimension which compose an attributed graph to solve the community detection problem. Recently, a density-based clustering model TCSC \cite{BES14} is proposed to detect clusters considering the connections in the network and the vertex attributes. 

Text information plays an important role in many heterogeneous network studies. Deng et al. \cite{DHZYL11} introduce a topic model with biased propagation to incorporate heterogeneous information network with topic modeling in a unified way. Furthermore, they \cite{DZH11} propose a joint probabilistic topic model for simultaneously modeling the contents of multi-typed objects of a heterogeneous information network. LSA-PTM \cite{WPJL13} is introduced to identify clusters of multi-typed objects by propagating the topics obtained by LSA on the HIN via the links between different objects. Incorporating both the document content and various links in the text related heterogeneous network, Wang et al. \cite{WPWYLH15} propose a unified topic model for topic mining and multiple objects clustering. Recently, CHINC \cite{WSERZH15} uses general-purpose knowledge as indirect supervision to improve the clustering results. 

User guide information is also integrated into clustering analysis. Sun et al. \cite{SNHYYY12} present a semi-supervised clustering algorithm to generate different clustering results with path selection according to user guidance. Luo et al. \cite{LPW14semi} firstly introduce the concept of relation-path to measure the similarity between same-typed objects and use the labeled information to weight relation-paths, and then propose SemiRPClus for semi-supervised learning in HIN. 

Clustering is usually an independent data mining task. However, it can be integrated with other mining tasks to improve performances through mutual enhancing. Recently, ranking-based clustering on heterogeneous information network has emerged, which shows its advantages on the mutual promotion of clustering and ranking. RankClus \cite{SHZYCW09} generates clusters for a specified type of objects in a bipartite network based on the idea that the qualities of clustering and ranking are mutually enhanced. The following work NetClus \cite{SYH09} is proposed to handle a network with the star-schema. Wang et al. \cite{WSYW13} introduce ComClus to promote clustering and ranking performance by applying star schema network with self loop to combine the heterogeneous and homogeneous information. In addition, a general method HeProjI is proposed to do ranking based clustering in heterogeneous networks with arbitrary schema by projecting the network into a sequence of sub-networks \cite{SWLYW14}. And Chen et al. \cite{CDSD15} propose a probabilistic generative model to simultaneously achieve clustering and ranking on a heterogeneous network with arbitrary schema. To make use of both textual information and heterogeneous linked entities, Wang et al. \cite{WDLDJH13} develop a clustering and ranking algorithm to automatically construct multi-typed topical hierarchies. What's more, Qiu et al. \cite{QCWL15} propose an algorithm OcdRank to combine overlapping community detection and community-member ranking together in directed heterogeneous social networks. 

Outlier detection is the process of finding data objects with behaviors that are very different from expectation. Outlier detection and clustering analysis are two highly related, but different-aimed tasks. To detect outliers, Gupta et al. \cite{GGH13} propose an outlier-aware approach based on joint non-negative matrix factorization to discover popular community distribution patterns. Furthermore, they propose to detect association-based clique outliers in heterogeneous networks given a conjunctive select query \cite{GGYCH13}. What's more, Zhuang et al. \cite{ZZBTCYH14} propose an outlier detection algorithm to find subnetwork outliers according to different queries and semantics. Also based on queries, Kuck et al. \cite{KZYCH15} propose a meta-path based outlierness measure for mining outliers in heterogeneous networks. 

In addition, some other information is also integrated. For example, a social influence based clustering framework SI-Cluster is proposed to analyze heterogeneous information networks based on both people's connections and their social activities \cite{ZL13}. Besides the traditional models employed in clustering on HIN, like topic model and spectral clustering, Alqadah et al. \cite{AB11} propose a novel game theoretic framework for defining and mining clusters in heterogeneous information networks.

\subsection{Classification}

Classification is a data analysis task where a model or classifier is constructed to predict class (categorical) labels. Traditional machine learning has focused on the classification of identically-structured objects satisfying independent identically distribution (IID). However, links exist among objects in many real-world datasets, which makes objects not satisfy IID. So link based object classification has received considerable attention, where a data graph is composed of a set of objects connected to each other via a set of links. Many methods extend traditional classification methods to consider correlations among objects \cite{CDI98, LMP01}. The link based object classification usually considers that objects and links in the graph are identical respectively. That is, the objects and links among them constitute a homogeneous network.  


\begin{figure}[htbp]
	\centering
	\includegraphics[width=11cm]{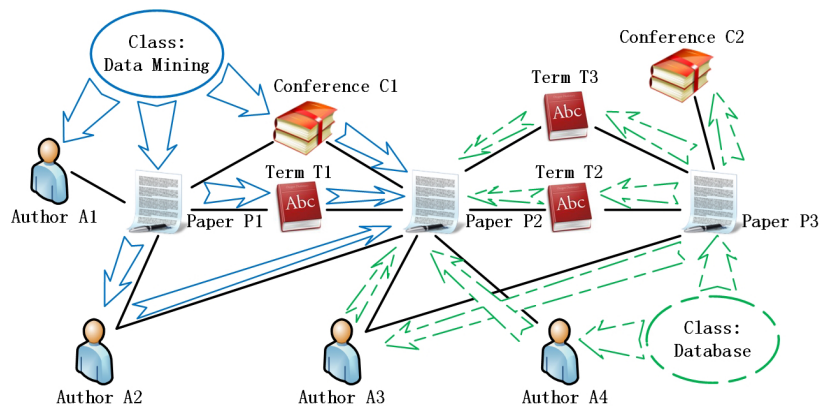}
	\caption{Classification process on a bibliographic heterogeneous network \cite{JSDHG10}.}
	\label{fig:transclass}
\end{figure}

Different from traditional classification researches, the classification problems studied in HIN have some new characteristics. First, the objects contained in HIN are different-typed, which means we can classify multiple types of objects simultaneously. Second, label knowledge can spread through various links among different-typed objects. Taking Fig. \ref{fig:transclass} as an example \cite{JSDHG10}, four types of objects (paper, author, conference and term) are interconnected by multi-typed links. The classification process can be intuitively viewed as a process of knowledge propagation throughout the network, where the arrows indicate possible knowledge flow. In this HIN condition, the label of objects is decided by the effects of different-typed objects along different-typed links.  

Many works extend traditional classification to heterogeneous information networks. Some works extend transductive classification task, which is to predict labels for the given unlabeled data. For example, GNetMine \cite{JSDHG10} is proposed to model the link structure in information networks with arbitrary network schema and arbitrary number of object/link types. Recently, Luo et al. propose HetPathMine \cite{LGWL14} to cluster with small labeled data on HIN through a novel meta path selection model, and Jacob et al. \cite{JDG14} propose a method to label nodes of different types by computing a latent representation of nodes in a space where two connected nodes tend to have close latent representations. Some works also extend inductive classification that is to construct a decision function in the whole data space. For example, Rossi et al. \cite{RFLR12} use a bipartite heterogeneous network to represent textual document collections and propose IMBHN algorithm to induce a classification model assigning weights to textual terms. 

Multi-label classification is prevalent in many real-world applications, where each example can be associated with a set of multiple labels simultaneously \cite{KCY13}. This kind of classification tasks are also extended to HIN. Angelova et al. \cite{AKW12} introduce a multi-label graph-based classification model for labeling heterogeneous networks by modeling the mutual influence between nodes as a random walk process. Kong et al. \cite{KCY13} use multiple types of relationships mined from the linkage structure of HIN to facilitate the multi-label classification process. Zhou et al. \cite{ZL14} propose an edge-centric multi-label classification approach considering both the structure affinity and the label vicinity. 

As a unique characteristic, meta path is widely used in classification on HIN. Meta paths are usually used for feature generation in many methods, such as GNetMine \cite{JSDHG10} and HetPathMine \cite{LGWL14}. Moreover, Kong et al. \cite{KYDW12} introduce the concept of meta-path based dependencies among objects to study the collective classification problem. 

Similar to clustering problem, classification is also integrated with other data mining tasks on HIN. Ranking-based classification is to integrate classification and ranking in a simultaneous, mutually enhancing process. Ji et al. \cite{KYDW12} propose a ranking-based classification framework, RankClass, to perform more accurate analysis. As an extension of RankClass, Chen et al. \cite{CCHM13} propose the F-RankClass for a unified classification framework that can be applied to binary or multi-class classification of unimodal or multimodal data. Some methods also integrate classification with information propagation. For example, Jendoubi et al. \cite{JMLY14} classify the social message based on its spreading in the network and the theory of belief functions.

\subsection{Link prediction}

Link prediction is a fundamental problem in link mining that attempts to estimate the likelihood of the existence of a link between two nodes, based on observed links and the attributes of nodes. Link prediction is often viewed as a simple binary classification problem: for any two potentially linked objects, predict whether the link exists (1) or not (0). One kind of approach is to make this prediction entirely based on structural properties of the network. Liben-Nowell and Kleinberg \cite{LK07} present a survey of predictors based on different graph proximity measures. Another kind of approach is to make use of attribute information for link prediction. For example, Popescul et al. \cite{PU03} introduce a structured logistic regression model that can make use of relational features to predict the existence of links. 


\begin{figure}[htbp]
	\centering
	\includegraphics[width=12cm]{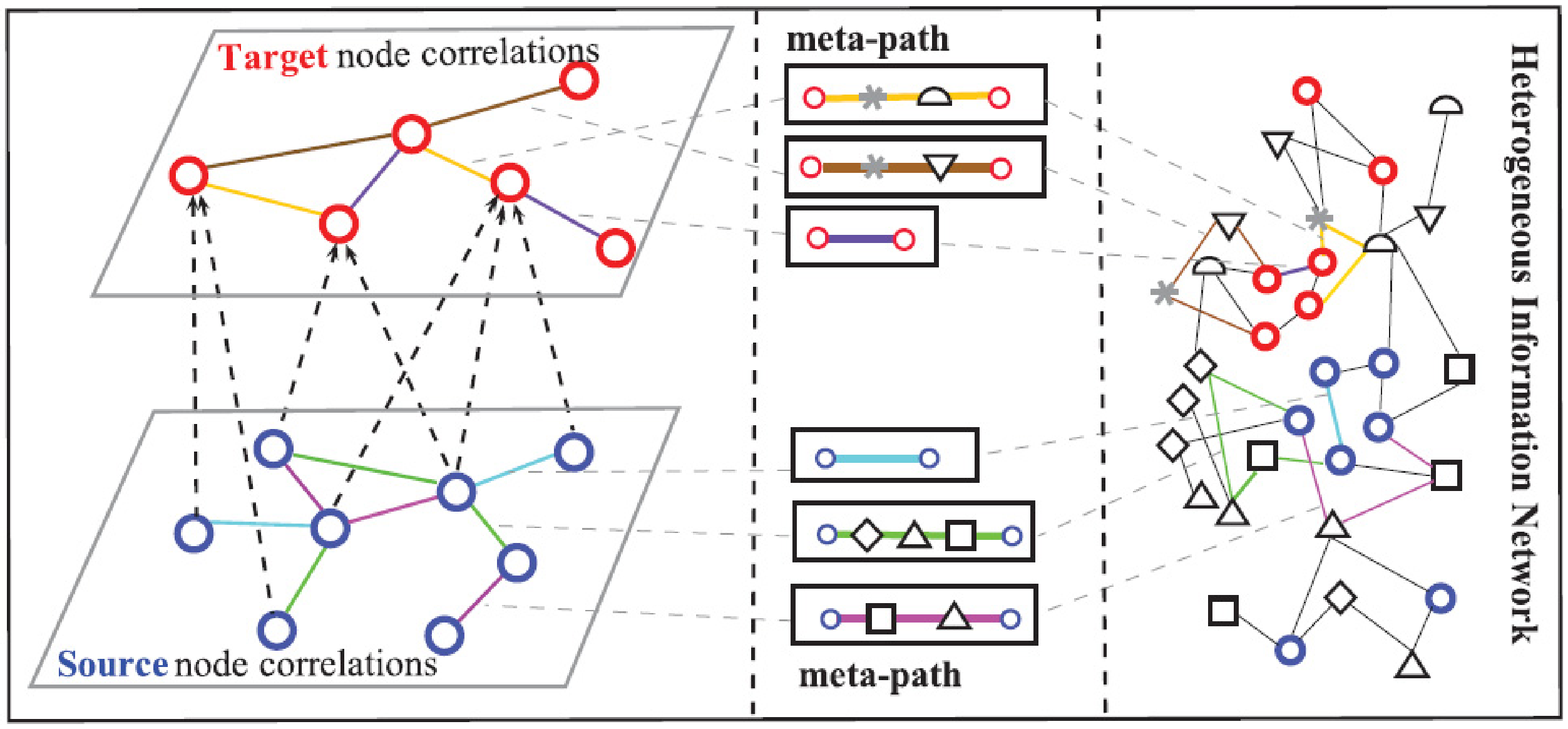}
	\caption{Collective link prediction in heterogeneous information network \cite{CKY14}.}
	\label{fig:colink}
\end{figure}

Link prediction in an HIN has been an important research topic for recent years, which has the following characteristics. First, the links to be predicted are of different types, since objects in HIN are connected with different types of links. Second, there are dependencies existing among multiple types of links. So link prediction in an HIN needs to collectively predict multiple types of links by capturing the diverse and complex relationships among different types of links and leveraging the complementary prediction information. For example, Fig. \ref{fig:colink} shows a collective prediction problem of multiple types of links in HIN \cite{CKY14}. Based on different meta paths, two objects have different link relations, and these relations have mutual effects.

Utilizing the meta path, many works employ a two-step process to solve link prediction problem in HIN. The first step is to extract meta path based feature vectors, while the second step is to train a regression or classification model to compute the existence probability of a link \cite{SBGAH11,YGZH12,CGWL13,CKY14,SHAC12}. For example, Sun et al. \cite{SBGAH11} propose PathPredict to solve the problem of co-author relationship prediction through meta path based feature extraction and logistic regression-based model. Zhang et al. \cite{ZYL15} use meta path based features to predict organization chart or management hierarchy. Utilizing diverse and complex linkage information, Cao et al. \cite{CKY14} design a relatedness measure to construct the feature vectors of links and propose an iterative framework to predict multiple types of links collectively. In addition, Sun et al. \cite{SHAC12} model the distribution of relationship building time with the use of the extracted topological features to predict when a certain relationship will be formed. 
  
Probabilistic models are also widely applied for link prediction tasks in HIN. Yang et al. \cite{YCSH12} propose a probabilistic method MRIP which models the influence propagating between heterogeneous relationships to predict links in multi-relational heterogeneous networks. Also, the TFGM model \cite{YTKZ12} defines a latent topic layer to bridge multiple networks and designs a semi-supervised learning model to mine competitive relationships across heterogeneous networks. Dong et al. \cite{DTWT12} develop a transfer-based ranking factor graph model that combines several social patterns with network structure information for link prediction and recommendation. Matrix factorization is another common tool to handle link prediction problems. For example, Huang et al. \cite{HNHT12} develop the joint manifold factorization (JMF) method to perform trust prediction with the ancillary rating matrix via aggregating heterogeneous social networks. 

The approaches mentioned above mainly focus on link prediction on one single heterogeneous network. Recently, Zhang et al. \cite{KZY13,ZKY13,ZKY14} propose the problem of link prediction across multiple aligned heterogeneous networks. A two-phase link prediction method is put forward in \cite{KZY13}. The first phase is to extract heterogeneous features from multiple networks, while the second phase is to infer anchor links by formulating it as a stable matching problem. In addition, Zhang et al. \cite{ZKY13} propose SCAN-PS to solve the social link prediction problem for new users using the ``anchors''. Furthermore, they propose the TRAIL \cite{ZKY14} method to predict social links and location links simultaneously. Also aimed at the cold start problem of new users, Liu et al. \cite{LX15} propose the aligned factor graph model for user-user link prediction problem by utilizing information from another similar social network. In order to  identify users from multiple heterogeneous social networks and integrate different networks, an energy-based model COSNET \cite{ZTYPY15} is proposed by considering both local and global consistency among multiple networks.

Most of the available works on link prediction are designed for static networks, however, the problem of dynamic link prediction is also very important and challenging. Taking into account both the dynamic and heterogeneous nature of web data, Zhao et al. \cite{ZBZY08} propose a general framework to characterize and predict community members from the evolution of heterogeneous web data. In order to solve the problem of dynamic link inference in temporal and heterogeneous information networks, Aggarwal et al. \cite{AXY12,AXY14} develop a two-level scheme which makes efficient macro- and micro-decisions for combining the topology and type information.

\subsection{Ranking}

Ranking is an important data mining task in network analysis, which evaluates object importance or popularity based on some ranking functions. Many ranking methods have been proposed in homogeneous networks. For example, PageRank \cite{PBMW98} evaluates the importance of objects through a random walk process, and HITS \cite{K99} ranks objects using the authority and hub scores. These approaches only consider the same type of objects in homogeneous networks. 

\begin{figure}[htbp]
	\centering
	\includegraphics[width=5cm]{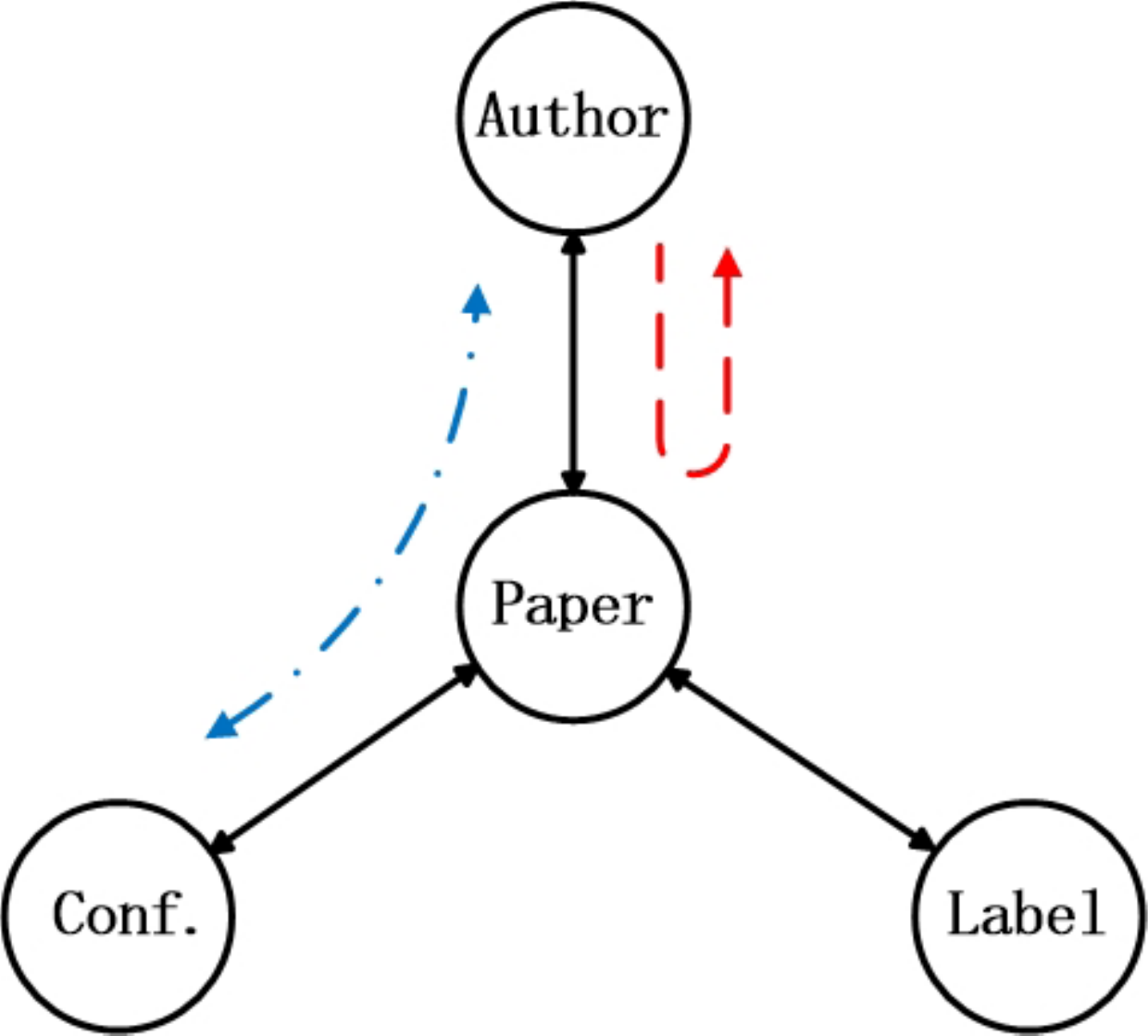}
	\caption{An example of ranking in bibliographic heterogeneous network \cite{LSPC14}.}
	\label{fig:hrank}
\end{figure}

Ranking in heterogeneous information networks is an important and meaningful task, but faces several challenges. First, there are different types of objects and relations in HIN, and treating all objects equally will mix different types of objects together. Second, different types of objects and relations in HIN carry different semantic meanings, which may lead to different ranking results. Taking the bibliographic heterogeneous network in Fig. \ref{fig:hrank} as an example, ranking on authors may have different results under different meta paths \cite{LSPC14}, since these meta paths will construct different link structures among authors. Moreover, the rankings of different-typed objects have mutual effects. For example, reputable authors usually publish papers on top conferences.  

The co-ranking problem on bipartite graphs has been widely explored in the past decades. For example, Zhou et al. \cite{ZOZG07} co-rank authors and their publications by coupling two random walk processes, and co-HITS \cite{DLK09} incorporates the bipartite graph with the content information and the constraints of relevance. Soulier et al. \cite{SJTB13} propose a bi-type entity ranking algorithm to rank jointly documents and authors in a bibliographic network regarding a topical query by combining content-based and network-based features. There are also some ranking works on multi-relational network. For example, MultiRank \cite{NLY11} is proposed to determine the importance of both objects and relations simultaneously for multi-relational data, and HAR \cite{LNY12} is proposed to determine hub and authority scores of objects and relevance scores of relations in multi-relational data for query search. These two methods focus on the same type of objects with multi-relations. Recently, Huang et al. \cite{HZJDWLAHLHV12} integrate both formal genre and inferred social networks with tweet networks to rank tweets. Although this work makes use of various types of objects in heterogeneous networks, it still ranks one type of objects. 

Considering the characteristics of meta path on HIN, some works propose path based ranking methods. For example, Liu et al. \cite{LYGS14} develop a publication ranking method with pseudo relevance feedback by leveraging a number of meta paths on the heterogeneous bibliographic graph. Applying the tensor analysis, Li et al. \cite{LSPC14} propose HRank to simultaneously evaluate the importance of multiple types of objects and meta paths. 

Ranking problem is also extended to HIN constructed by social media network. For image search in social media, Tsai et al. \cite{TAH14} propose SocialRank which uses social hints for image search and ranking in social networks. To identify high quality objects (questions, answers, and users) in Q\&A systems, Zhang et al. \cite{ZKJCP14} devise an unsupervised heterogeneous network based framework to co-rank multiple objects in Q\&A sites. For heterogeneous cross-domain ranking problem, Wang et al. \cite{WTFCTY13} propose a general regularized framework to discover a latent space for two domains and minimize two weighted ranking functions simultaneously in the latent space. Considering the dynamic nature of literature networks, a mutual reinforcement ranking framework is proposed to rank the future popularity of new publications and young researchers simultaneously \cite{WXZLYS14}.

\subsection{Recommendation}

Recommender systems help consumers to make product recommendations that are likely to be of interest to the user such as books, movies, and restaurants. It uses a broad range of techniques from information retrieval, statistics, and machine learning to search for similarities among items and customer preferences. Traditional recommender systems normally only utilize the user-item rating feedback information for recommendation. Collaborative filtering is one of the most popular techniques, which includes two types of approaches: memory-based methods and model-based methods. Recently, matrix factorization has shown its effectiveness and efficiency in recommender systems, which factorizes the user-item rating matrix into two low rank user-specific and item-specific matrices, and then utilizes the factorized matrices to make further predictions \cite{SJ03}. With the prevalence of social media, more and more researchers study social recommender system, which utilizes social relations among users \cite{MKL09,YSL12}.  

\begin{figure}[htbp]
	\centering
	\includegraphics[width=9cm]{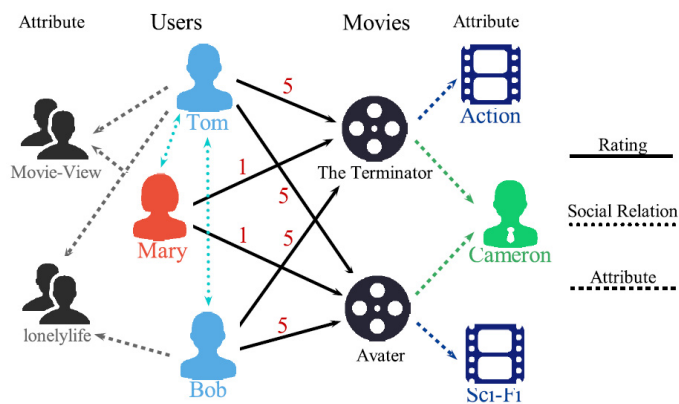}
	\caption{An example of heterogeneous information network for movie recommendation \cite{SZLYYW15}.}
	\label{fig:semrec}
\end{figure}

Recently, some researchers have begun to be aware of the importance of heterogeneous information for recommendations. The comprehensive information and rich semantics of HIN make it promising to generate better recommendations. Fig. \ref{fig:semrec} shows such an example in movie recommendation \cite{SZLYYW15}. The HIN not only contains different types of objects (e.g., users and movies) but also illustrates all kinds of relations among objects, such as viewing information, social relations, and attribute information. Constructing heterogeneous networks for recommendation can effectively fuse all kinds of information, which can be potentially utilized for recommendation. Moreover, the objects and relations in the networks have different semantics, which can be explored to reveal subtle relations among objects.

Meta path is well used to explore the semantics and extract relations among objects. Shi et al. \cite{SZKY12} implement a semantic-based recommendation system HeteRecom, which employs the semantics information of meta path to evaluate the similarities between movies. Furthermore, considering the attribute values, such as rating score on links, they model the recommender system as a weighted HIN and propose a semantic path based personalized recommendation method SemRec \cite{SZLYYW15}. In order to take full advantage of the relationship heterogeneity, Yu et al. \cite{YRSSKGNH13,YRSGSKNH14} introduce meta-path-based latent features to represent the connectivity between users and items along different types of paths, and then define recommendation models at both global and personalized levels with Bayesian ranking optimization techniques. Also based on meta path, Burke et al. \cite{BVM14} present an approach for recommendation which incorporates multiple relations in a weighted hybrid. 

A number of approaches employ heterogeneous information network to fuse various kinds of information. Utilizing different contexts information, Jamali et al. \cite{JL13} propose a context-dependent matrix factorization model which considers a general latent factor for every entity and context-dependent latent factors for every context. Using user implicit feedback data, Yu et al. \cite{YRSSKGNH13,YRSGSKNH14} solve the global and personalized entity recommendation problem. Based on related interest groups, Ren et al. \cite{RLYKGWH14} propose a cluster-based citation recommendation framework to predict each query's citations in bibliographic networks. Similarly, Wu et al. \cite{WCYHW15} exploit graph summarization and content-based clustering for media recommendation with the interest group information. Based on multiple heterogeneous network features, Yang et al. \cite{YSMZWH15} model multiple features into a unified framework with a SVM-Rank based method. In addition, using multiple types of relations, Luo et al. \cite{LPW14} propose a social collaborative filtering algorithm.

\subsection{Information Fusion}

\begin{figure}[htbp]
	\centering
	\subfigure[Network alignment \cite{FF11}]{
		\label{fig:Network alignment}
		\begin{minipage}[t]{0.35\textwidth}
			\includegraphics[width=6cm]{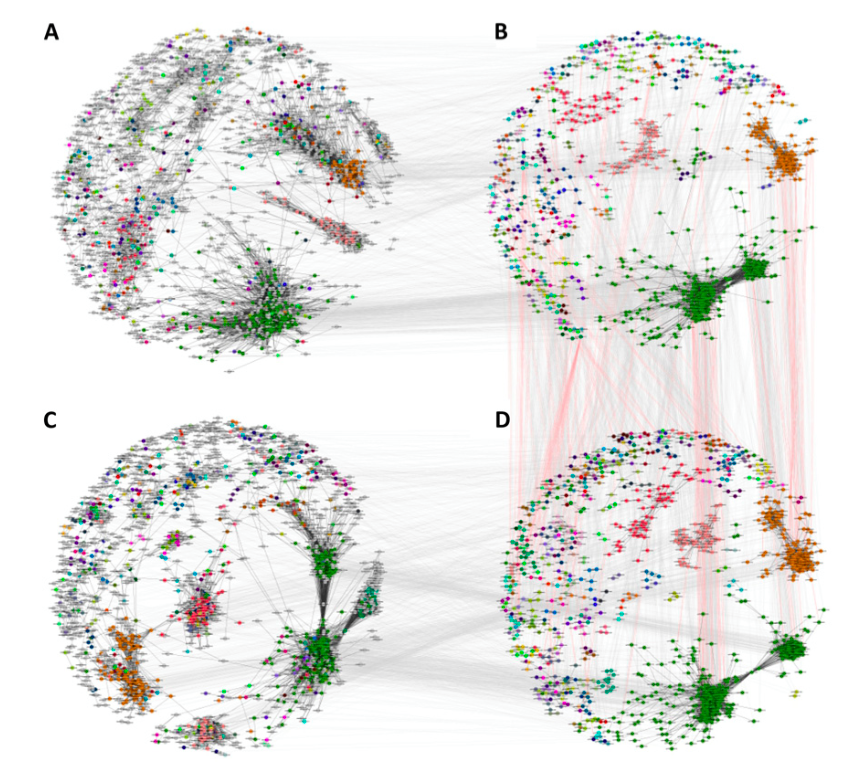}
		\end{minipage}
	}
	\hspace{25pt}
	\subfigure[Subgraph isomorphism \cite{KGPH12}]{
		\label{fig:Subgraph isomorphism}
		\begin{minipage}[t]{0.4\textwidth}
			\includegraphics[width=7cm]{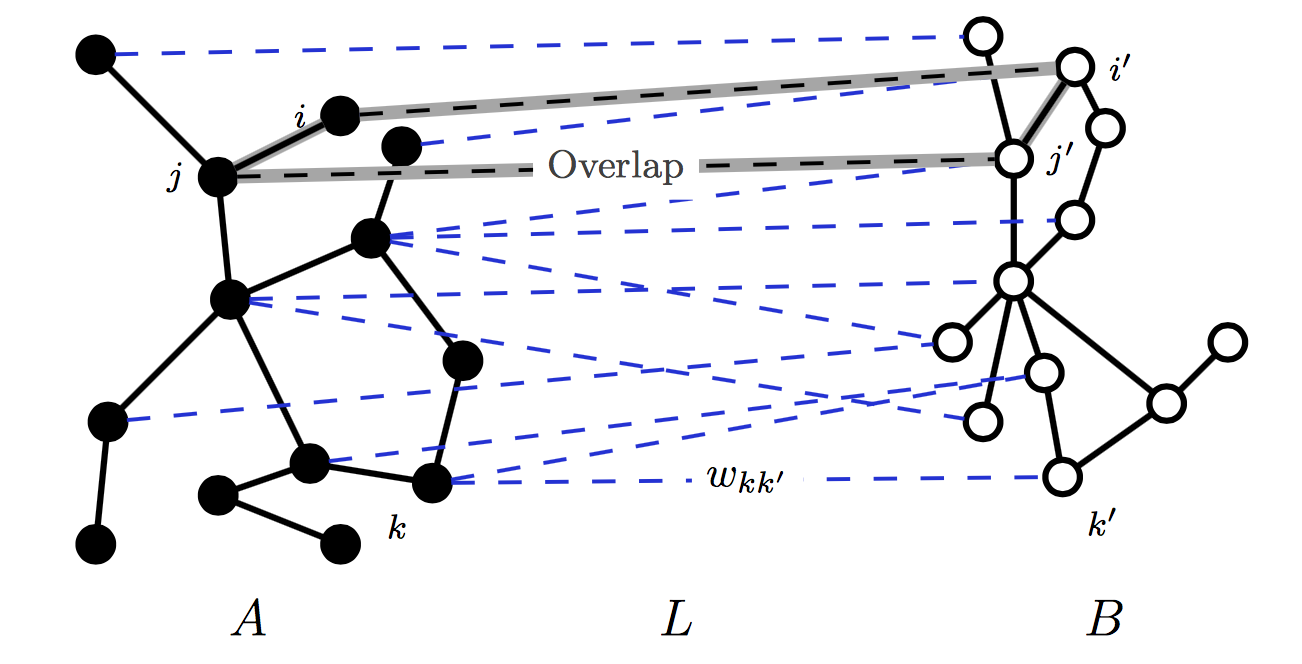}
		\end{minipage}
	}
	\caption{An example of information network alignment.}
	\label{fig:alignment}
\end{figure}

Information fusion denotes the process of merging information from heterogeneous sources with differing conceptual, contextual and typographical representations. Due to the availability of various data sources, fusing these scattered distributed information sources has become an important research problem. In the past decades, dozens of papers have been published on this topic in many traditional data mining areas, e.g., data schemas integration in data warehouse \cite{MGR02}, protein-protein interaction (PPI) networks and gene regulatory networks matching in bioinformatics \cite{KBS08, SP12, LLBSB09, SXB07}, and ontology mapping in web semantics \cite{DMDH04}. Nowadays, with the surge of HIN, information fusion across multiple HINs has become a novel yet important research problem. By fusing information from different HINs, we can obtain a more comprehensive and consistent knowledge about the common information entities shared in different HINs, including their structures, properties, and activities.

To fuse the information in multiple HINs, an important prerequisite will be to align the HINs via the shared common information entities, which can be users in social networks, authors in bibliographical networks, and protein molecules in biological networks. For instance, Fig.~\ref{fig:apa} is about the alignment of two heterogeneous social networks via the shared users, and Fig.~\ref{fig:Network alignment} shows an example about alignment of two biological networks via molecules of common properties. Perfect HIN alignment is a challenging problem as the underlying subgraph isomorphism problem, as shown in Fig.~\ref{fig:Subgraph isomorphism}, is actually NP-complete \cite{K09}. Meanwhile, based on the structure and attribute information available in HINs, a large number of approximated HIN alignment algorithms have been proposed so far. Enlightened by the homogeneous network alignment method in \cite{U88}, Koutra et al. \cite{KTL13} propose to align two bipartite graphs with a fast network alignment algorithm. Zafarani et al. \cite{ZL13} propose to match users across social networks based on various node attributes, e.g., username, typing patterns and language patterns etc. Kong et al. \cite{KZY13} formulate the heterogeneous social network alignment problem as an anchor link prediction problem. A two-step supervised method MNA is proposed in \cite{KZY13} to infer potential anchor links across networks with heterogeneous information in the networks. However, social networks in the real world are actually mostly partially aligned and lots of users are not anchor users. Zhang et al. have proposed the partial network alignment methods based on supervised learning setting and PU learning setting in \cite{ZSWKY15} and \cite{ZY15-IJCAI} respectively. In addition to these pairwise social network alignment problems, multiple (more than two) social networks can be aligned simultaneously. Zhang et al. \cite{ZY15-ICDM} discover that the inferred cross-network mapping of entities in social network alignment should meet the transitivity law and has an inherent one-to-one constraint. A new multiple social alignment framework is introduced in \cite{ZY15-ICDM} to minimize the alignment costs and preserve the transitivity law and one-to-one constraint on the inferred mappings.

By fusing multiple HINs, the heterogeneous information available in each network can be transferred to other aligned networks and lots of application problems on HIN, e.g., link prediction and friend recommendation \cite{ZYZ14, ZY15-IJCAI}, community detection \cite{ZY15-SDM}, information diffusion \cite{ZZWPX15}, will benefit from it a lot. 

Via the inferred mappings, Zhang et al. propose to transfer heterogeneous links across aligned networks to improve quality of predicted links/recommended friends \cite{ZYZ14, ZY15-IJCAI}. For new networks \cite{ZKY14} and new users \cite{ZKY13} with little social activity information, the transferred information can greatly overcome the cold start problem when predicting links for them. What's more, information about the shared entities across aligned networks can provide us with a more comprehensive knowledge about the community structures formed by them. By utilizing the information across multiple aligned networks, Zhang et al. \cite{ZY15-Big} propose a new model to refine the clustering results of the shared entities with information in other aligned networks mutually. Jin et al. \cite{JZYYL14} propose a scalable framework to study the synergistic partitioning of multiple aligned large-scale networks, which takes the relationships among different networks into consideration and tries to maintain the consistency on partitioning the same nodes of different networks into the same partitions. Zhang et al. \cite{ZY15-SDM} study the community detection in emerging networks with information transferred from other aligned networks to overcome the cold start problem. In addition, by fusing multiple heterogeneous social networks, users in networks will be extensively connected with each other via both intra-network connections (e.g., friendship connections among users) and inter-network connections (i.e., the inferred mappings across networks). As a result, information can reach more users and achieve broader influence across the aligned social networks. Zhan et al. propose a new model to study the information diffusion process across multiple aligned networks in \cite{ZZWPX15}.

\subsection{Other applications}

Besides the tasks discussed above, there are many other applications in heterogeneous networks, such as influence propagation and privacy risk problem.
To quantitatively learn influence from heterogeneous networks, Liu et al. \cite{LTHY12} first use a generative graphical model to learn the direct influence, and then use propagation methods to mine indirect and global influence. Using meta paths, Zhan et al. \cite{ZZWPX15} propose a model M\&M to solve the influence maximization problem in multiple partially aligned heterogeneous online social networks. For privacy risk in anonymized HIN, Zhang et al. \cite{ZXCGHW14} present a de-anonymization attack that exploits the identified vulnerability to prey upon the risk. Aiming at the inferior performances of unsupervised text embedding methods, Tang et al. \cite{TQM15} propose a semi-supervised representation learning method for text data, in which labeled information and different levels of word co-occurrence information are represented as a large-scale heterogeneous text network.  

\subsection{Application Systems}
Besides many data mining tasks explored on heterogeneous network, some demo systems have designed prototype applications on HIN. Through employing a path-based relevance measure to evaluate the relevance between any-typed objects and capture the subtle semantic containing in meta path, Shi et al. \cite{SZKY12} implement a HeteRecom system for semantic recommendation. Yu et al. \cite{YSZH12} demonstrate a prototype system on query-driven discovery of semantically similar substructures in heterogeneous networks. Danilevsky et al. \cite{DWTNCDWH13} present the AMETHYST system for exploring and analyzing a topical hierarchy constructed from an HIN. In LikeMiner system, Jin et al. \cite{JWLYH11} introduce a heterogeneous network model for social media with `likes', and propose `like' mining algorithms to estimate representativeness and influence of objects. Meanwhile, they design SocialSpamGuard \cite{JLLH11}, a scalable and online social media spam detection system for social network security. Taking DBLP as an example, Tao et al. \cite{TYLBCHKSWW13} construct a Research-Insight system to demonstrate the power of database-oriented information network analysis including ranking, clustering, classification, recommendation, and prediction. Furthermore, they construct a semi-structured news information network NewsNet and develop a NewsNetExplorer system \cite{TBHJWNELRS14} to provide a set of news information network exploration and mining functions. 

Some real application systems also have been designed. One of the most famous works is ArnetMiner \footnote{http://aminer.org/} \cite{TZYLZS08}, which offers comprehensive search and mining services for academic community. ArnetMiner not only provides abundant online academic services but also offers ideal test platform for heterogenous information network analysis. PatentMiner \footnote{http://pminer.org/home.do?m=home} \cite{TWYHZYGHXL12} is another application which is a general topic-driven framework for analyzing and mining heterogeneous patent networks.

\section{Advanced Topics}

Although many data mining tasks have been exploited in heterogeneous information network, it is still a young and promising research field. Here we illustrate some advanced topics, including challenging research issues and unexplored tasks, and point out some potential future research directions. 

\subsection{More complex network construction}
There is a basic assumption in contemporary  researches that a heterogeneous information network to be investigated is well-defined, and objects and links in the network are clean and unambiguous. However, it is not the case in real applications. In fact, constructing heterogeneous information network from real data often faces challenges.  

If the networked data are structured data, like relational database, it may be easy to construct a heterogeneous information network with well-defined schema, such as DBLP network \cite{SHYYW11} and Movie network \cite{SZKY12, YRSSKGNH13}. However, even in this kind of heterogeneous network, objects and links can still be noisy. (1) Objects in a network may not exactly correspond to entities in real world, such as duplication of name \cite{YHY07} in bibliography data. That is, one object in a network may refer to multiple entities, or different objects may refer to the same entity. We can integrate entity resolution \cite{BG07} with network mining to clean objects or links beforehand. For example, Shen et al. \cite{SHW14} propose a probabilistic model SHINE to link named entity mentions detected from the unstructured Web text with their corresponding entities existing in a heterogeneous information network. Ren et al. \cite{RKWTVH15} propose a relation phrase-based entity recognition framework, called ClusType. The framework runs data-driven phrase mining to generate entity mention candidates and relation phrases, and enforces the principle that relation phrases should be softly clustered when propagating type information in a heterogeneous network constructed by argument entities. (2) Relations among objects may not be explicitly given or not complete sometimes, e.g., the advisor-advisee relationship in the DBLP network \cite{WHJTZYG10}. Link prediction \cite{LK07} can be employed to fill out the missing relations for comprehensive networks. (3) Objects and links may not be reliable or trustable, e.g., the inaccurate item information in an E-commerce website and conflicting information of certain objects from multiple websites. So it is the first step to clean and integrate networked data for high-quality network construction, such as trustworthiness modeling \cite{ZRGH12,YHY08} and spam detection \cite{WXLY12}. 

If the networked data are unstructured data, such as text data, multimedia data and multi-lingual data, it becomes more challenging to construct qualified heterogeneous information networks. In order to construct high-quality HINs, information extraction, natural language processing, and many other techniques should be integrated with network construction. Mining quality phrases is a critical step to form entities of networks from text data. Kishky et al. \cite{KSWVH14} propose a computationally efficient and effective model ToPMine, which first executes  a phrase mining framework to segment a document into single and multi-word phrases, and then employs a new topic model that operates on the induced document partition. Furthermore, Liu et al. \cite{LSWRH15} propose an effective and scalable method SegPhrase+ that integrates quality phrases extraction with phrasal segmentation. Relationship extraction is another important step to form links among objects in network. Wang et al. \cite{WHJTZYG10} mine hidden advisor-advisee relationships from bibliographic data, and they further infer hierarchical relationships among partially ordered objects with heterogeneous attributes and links \cite{WHLL12}. Broadly speaking, we can also extract entity and relationship to construct heterogeneous network from multimedia data and multi-lingual data, as we have done on text data.

\subsection{More powerful mining methods}
For ubiquitous heterogeneous information networks, numbers of mining methods have been proposed on many data mining tasks. As we have said, heterogeneous information networks have two important characteristics: complex structure and rich semantics. According to these two characteristics, we summarize the contemporary works and point out future directions. 

\subsubsection{Network Structure}
In heterogeneous network, objects can be organized in different forms. Bipartite graph is widely used to organize two types of objects and the relations among them \cite{JL13, LZY05, SHZYCW09}. As an extension of bipartite graphs, $K$-partite graphs \cite{LWZY06} are able to represent multiple types of objects. Recently, heterogeneous networks are usually organized as star-schema networks, such as bibliographic data \cite{SYH09,SHYYW11,SKYX12} and movie data \cite{SZKY12, YRSSKGNH13}. To combine the heterogeneous and homogeneous information, star schema with self loop is also proposed \cite{WSYW13}. Different from only one hub object type existing in star schema network, some networked data have multiple hub object types, e.g., the bioinformatics data \cite{SWLYW14}. For this kind of networks, Shi et al. \cite{SWLYW14} propose a HeProjI method which projects a general heterogeneous network into a sequence of sub-networks with bipartite or star-schema structure.

In real applications, the networked data are usually more complex and irregular. Some real networks may contain attribute values on links, and these attribute values may contain important information. For example, users usually rate movies with a score from 1 to 5 in movie recommender system, where the rating scores represent users' attitudes to movies, and the ``author of'' relation between authors and papers in bibliographic networks can take values (e.g., 1, 2, 3) which means the order of authors in the paper. In this kind of applications, we need to consider the effect of attribute values on the weighted heterogeneous information network \cite{SZLYYW15}.  There are some time-series data, for example, a period of biographic data and rating information of users and movies. For this kind of data, we need to construct dynamic heterogeneous network \cite{STHGZ10} and consider the effect of time factor. In some applications, one kind of objects may exist in multiple heterogeneous networks \cite{KZY13,ZKY13}. For example, users usually co-exist in multiple social networks, such as Facebook, Google+, and Twitter. In this kind of applications, we need to align users in different networks and effectively fuse information from different networks \cite{ZSWKY15,ZY15-IJCAI,ZY15-ICDM}. More broadly, many networked data are difficult to be modeled with heterogeneous network with a simple network schema. For example, in RDF data, there are so many types of objects and relations, which cannot be described with network schema \cite{MCMSZ15,WSERZH15}. Many research problems arise with this kind of schema-rich HINs, for example, management of objects and relations with so many types and automatic generation of meta paths. As the real networked data become more complex, we need to design more powerful and flexible heterogeneous networks, which also provides more challenges for data mining. 

\subsubsection{Semantic Mining}
As a unique characteristic, objects and links in HIN contain rich semantics. Meta path can effectively capture subtle semantics among objects, and many works have exploited the meta path based mining tasks. For example, in similarity measure task, object pairs have different similarities under different meta paths \cite{SHYYW11,SKYX12}; in recommendation task, different items will be recommended under different paths \cite{SZLYYW15}.  In addition, meta path is also widely used for feature extraction. Object similarity can be measured under different meta paths, which can be used as feature vectors for many tasks, such as clustering \cite{SNHYYY12}, link prediction \cite{CKY14}, and recommendation \cite{YRSGSKNH14}.  

However, some researchers have noticed the shortcomings of meta path. In some applications, meta path fails to capture more subtle semantics. For example, the ``Author-Paper-Author" path  describes the collaboration relation among authors. However, it cannot depict the fact that Philip S. Yu and Jiawei Han have many collaborations in data mining field but they seldom collaborate in information retrieval field. In order to overcome the shortcoming existing in meta path, Shi et al. \cite{LSPC14} propose the constrained meta path concept, which can confine some constraints on objects. Taking Fig. \ref{fig:dblp} as an example, the constrained meta path $APA|P.L=''Data\ Mining''$ represents the co-author relation of authors in data mining field through constraining the label of papers with ``Data Mining''. Moreover, Liu et al. \cite{LYGS14} propose the concept ``restricted meta-path" which enables in-depth knowledge mining on the heterogeneous bibliographic networks by allowing restrictions on the node set. In addition, traditional HIN and meta path do not consider the attribute values on links, while weighted links are very common in real applications. Examples include rating scores between users and items in recommender system and the order of authors in papers in bibliographic network. Taking Fig. \ref{fig:semrec} as an example, the rating relation between users and movies can take scores from 1 to 5. Under the meta path ``User-Movie-User'', Tom has the same similarity with Mary and Bob, but we can find that they may have totally different tastes due to different rating scores. Shi et al. \cite{SZLYYW15} propose weighted meta path to consider attribute values on links and more subtly capture path semantics through distinguishing different link attribute values. As an effective semantic capture tool, meta path has shown its power in semantic capture and feature selection. However, it may be coarse in some applications, so we need to extend traditional meta path for more subtle semantic capture. Broadly speaking, we can also design new, and more powerful semantic capture tools. 

More importantly, the meta path approach faces challenges on path selection and their weight importances. How can we select meta paths in real applications? Theoretically, there are infinite meta paths in an HIN. In contemporary works, the network schema of HIN is usually small and simple, so we can assign some short and meaningful meta paths according to domain knowledge and experiences. Sun et al. \cite{SHYYW11} have validated that the long meta paths are not meaningful and they fail to produce good similarity measures. However, there is no work to study the effect of long meta paths on other mining tasks. In addition, there are so many meta paths even for short paths in some complex networks, like RDF network. It is a critical task to automatically extract meta paths in this condition. Recently, Meng et al. \cite{MCMSZ15} study how to discover meta paths automatically which can best explain the relationship between node pairs. Another important issue is to automatically determine the weights of meta paths. Some methods have been proposed to explore this issue. For example, Lao et al. \cite{LC10} employ a supervised method to learn weights, and Sun et al. \cite{SNHYYY12} combine meta-path selection and user-guided information for clustering. Some interesting works are still worth doing. The ideal path weights learned should embody the importance of paths and reflect users' preferences. However, the similarity evaluations based on different paths have significant bias, which may make path weights hard to reflect path importances. So prioritized path weights are needed. In addition, if there are numerous meta paths in real applications (e.g., RDF network), the path weight learning will be more important and challenging. 

\begin{figure}[htbp]
\begin{center}
	\includegraphics[width=15cm]{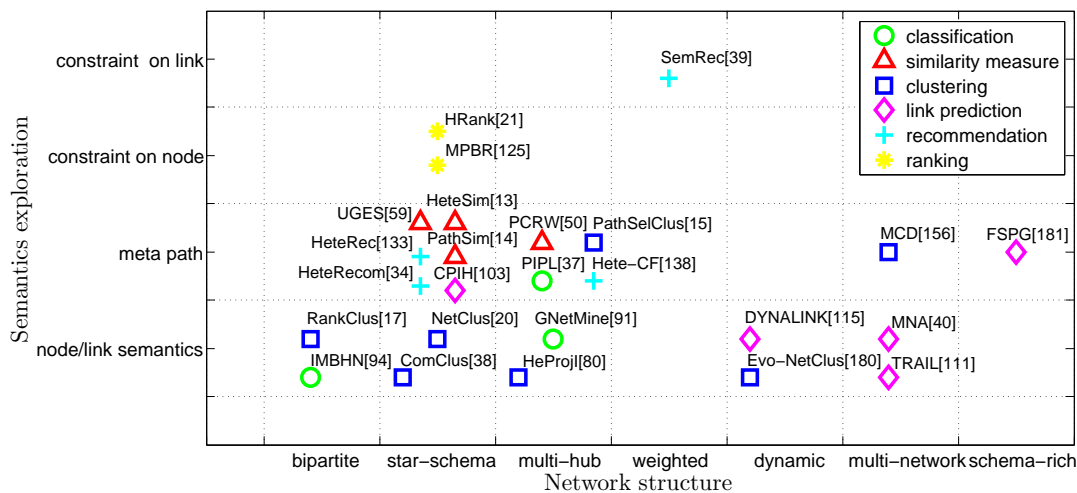}
	\caption{Summarization of typical works  on HIN according to network structure and semantic exploration.}
\end{center}
\label{fig:ss}
\end{figure}

In Fig. 12, we summarize some typical works in the HIN field from two perspectives: network structure and semantic exploration. We respectively select several typical works from six mining tasks mentioned above, and put these works in a coordinate according to network structure and semantics explored in these works. Note that we denominate those un-named methods with the first letter of keywords in the title, such as UGES \cite{YSNMH12} and CPIH \cite{YGZH12}. Along the X-axis, the network structure becomes more complex, and semantics information becomes richer along the Y-axis. For example, RankClus \cite{SHZYCW09} is designed for bipartite networks and only captures link semantics (different-typed links contain different semantics). While PathSim \cite{SHYYW11} can deal with more complex star-schema networks and use meta path to mine deeper semantics. Further, SemRec \cite{SZLYYW15} adds constraints to links to explore more subtle semantic information in a weighted HIN. From the figure, we can also find that most contemporary works focus on simple network structures (e.g., bipartite or star schema networks) and primary semantic exploration (e.g., meta path). In the future, we can exploit more complex heterogeneous networks with more powerful semantics capture tools. 

\subsection{Bigger networked data}
In order to illustrate the benefits of HIN, we need to design data mining algorithms on big networked data in wider domains. The variety is an important characteristic of big data. HIN is a powerful tool to handle the variety of big data, since it can flexibly and effectively integrate varied objects and heterogeneous information. However, it is non-trivial work to build a real HIN based analysis system. Besides research challenges mentioned above, such as network construction, it will face many practical technique challenges. Real HIN is huge, even dynamic, so it usually cannot be contained in memory and cannot be handled directly. We know that a user at a time could be only interested in a tiny portion of nodes, links, or sub-networks. Instead of directly mining the whole network, we can mine hidden but small networks ``extracted'' dynamically from some existing networks, based on user-specified constraints or expected node/link behaviors. How to discover such hidden networks and mine knowledge (e.g., clusters, behaviors, and anomalies) from such hidden but non-isolated networks could be an interesting but challenging problem.  

Most of contemporary data mining tasks on HIN only work on small dataset, and fail to consider the quick and parallel process on big data. Some research works have begun to consider the quick computation of mining algorithms on HIN. For example, Sun et al. \cite{SHYYW11} design a co-clustering based pruning strategy to fasten the processing speed of PathSim. Lao et al. \cite{LC10FAST} propose the quick computation strategies of PCRW, and Shi et al. \cite{SKHYW14, MSLZW14} also consider the quick/parallel computation of HeteSim. In addition, cloud computing also provides an option to handle big networked data. Although parallel graph mining algorithms \cite{C09} and platforms \cite{KTF09} have been proposed, parallel HIN analysis methods face some unique challenges. For example, the partition of HIN needs to consider the overload balances of computing nodes, as well as balances of different-typed nodes. Moreover, it is also challenging to mine integrated path semantics in partitioned subgraphs.

\subsection{More applications}
Due to unique characteristics of HIN, many data mining tasks have been explored on HIN, which are summarized as above.   In fact, more data mining tasks can be studied on HIN. Here we introduce two potential applications. 

The online analytical processing (OLAP) has shown its power in multidimensional analysis of structured, relational data \cite{C96}. The similar analysis can also be done, when we view a heterogeneous information network from different angles and at different levels of granularity. Taking a bibliographic network as an example, we can observe the change of published papers on a conference in the time or district dimension, when we designate papers and conferences as the object types and publish relations as the link type. Some preliminary studies have been done on this issue. Zhao et al. \cite{ZLXH11} introduce graph cube to support OLAP queries effectively on large multidimensional networks; Li et al. \cite{LYZXL11} design InfoNetOLAPer to provide topic-oriented, integrated, and multidimensional organizational solutions for information networks. Yin et al. \cite{YWZ12} have developed a novel HMGraph OLAP framework to mine multi-dimensional heterogeneous information networks with more dimensions and operations. These works consider link relation as a measure. However, they usually ignore semantic information in heterogeneous networks determined by multiple nodes and links. So the study of online analytical processing of heterogeneous information networks is still worth exploring. 

Information diffusion is a vast research domain and has attracted research interests from many fields, such as physics, biology, etc. Traditional information diffusion is studied on homogeneous networks \cite{GGLT04}, where information is propagated in one single channel. However, in many applications, pieces of information or diseases are propagated among different types of objects. For example, diseases could propagate among people, different kinds of animals and food, via different channels. Few works explore this issue. Liu et al. \cite{LTHJY10} propose a generative graphical model which utilizes the heterogeneous link information and the textual content associated with each node to mine topic level direct influence. In order to capture better spreading models that represent the real world patterns, it is desirable to pay more attention to the study of information diffusion in heterogeneous information networks.  

\section{Conclusion}
There is a surge on heterogenous information network analysis in recent years because of rich structural and semantic information in this kind of networks. This paper provides an extensive survey in this rapidly expanding field. We present the recent developments of different data mining tasks on heterogenous information network, along with future development directions. Hopefully, this survey will give researchers an understanding of the fundamental issues and a good starting point to work on this field.

%
%
%
%

\newpage

\bibliographystyle{IEEEtran}
\bibliography{IEEEabrv,References}




\end{document}